\documentclass[aps,pra,amsmath,superscriptaddress,onecolumn,longbibliography,nofootinbib]{revtex4-1}
\usepackage{graphicx}
\usepackage{caption}
\usepackage{subcaption}
\usepackage{amsmath}
\usepackage{amsfonts}
\usepackage{amssymb}
\usepackage{bbold}
\usepackage{algorithm}
\usepackage{algpseudocode}
\usepackage{color}
\usepackage{natbib}

\makeatletter

\newcommand{\Rmnum}[1]{\expandafter\@slowromancap\romannumeral #1@}
\makeatother

\begin{document}
\title{A Matrix Product State Representation of Boolean Functions }

\author{Umut Eren Usturali}
\affiliation{Physics Department, Boston University, Boston, Massachusetts 02215, USA}

\author{Claudio Chamon}
\affiliation{Physics Department, Boston University, Boston, Massachusetts 02215, USA}

\author{Andrei E. Ruckenstein}
\affiliation{Physics Department, Boston University, Boston, Massachusetts 02215, USA}

\author{Eduardo R. Mucciolo}
\email[Corresponding author: ] {eduardo.mucciolo@ucf.edu}
\affiliation{Department of Physics, University of Central Florida, Orlando, Florida 32816, USA}

\date{\today}

\begin{abstract}
We introduce a novel normal form representation of Boolean functions
in terms of products of binary matrices, hereafter referred to as the
Binary Matrix Product (BMP) representation. BMPs are analogous to the
Tensor-Trains (TT) and Matrix Product States (MPS) used, respectively,
in applied mathematics and in quantum many-body physics to accelerate
computations that are usually inaccessible by more traditional
approaches. BMPs turn out to be closely related to Binary Decision
Diagrams (BDDs), a powerful compressed representation of Boolean
functions invented in the late 80s by Bryant~\cite{Bryant1986} that has found a
broad range of applications in many areas of computer science and
engineering. We present a direct and natural translation of BMPs into
Binary Decision Diagrams (BDDs), and derive an elementary set of
operations used to manipulate and combine BMPs that are analogous to
those introduced by Bryant for BDDs. Both BDDs and BMPs are
practical tools when the complexity of these representations, as
measured by the maximum bond dimension of a BMP (or the accumulated bond
dimension across the BMP matrix train) and the number of nodes of a
BDD, remains polynomial in the number of bits, $n$. In both cases,
controlling the complexity hinges on optimizing the order of the
Boolean variables. BMPs offer the advantage that their construction
and manipulation rely on simple linear algebra -- a compelling feature
that can facilitate the development of open-source libraries that are
both more flexible and easier to use than those currently available
for BDDs. An initial implementation of a BMP library is available on
GitHub~\cite{BMP-library}, with the expectation that the close conceptual connection 
to tensor-train (TT) and matrix product state (MPS) techniques
-- widely employed in applied mathematics and quantum many-body
physics -- will motivate further development of BMP methods by
researchers in these fields, potentially enabling novel applications
to classical and quantum computing.

\end{abstract}

\maketitle

%%%%%%%%%%%%%%%%%%%%%%%%%%%%%%%%%%%%%%%%%%%%%%%%%%%%%%%%%%%%%%%%%%%%%%%%%
\section{Introduction}
\label{sec:intro}

In parallel to the intense activity and fierce competition focused on
practical demonstrations of quantum supremacy \cite{Arute2019,
  Zhu2022, king2024, Daley2022}, a new line of research is emerging
that employs quantum ideas and approaches to speed up and optimize
solutions to complex classical computational problems. These quantum
inspired approaches promise to significantly impact computational
sciences before the holy grail of quantum computing is likely to be
achieved at scale. This paper introduces a new representation of
Boolean functions that employs {\it Boolean Matrix Products} (BMP),
analogues of Matrix Product States (MPS) \cite{Perez-Garcia2007,
  Verstraete2008}, special forms of tensor network states
\cite{Orus2014,Orus2019}, developed and extensively studied since the
early 1990s as approximations of quantum states of complex many-body
quantum systems with small to moderate entanglement.

While we believe ours is the first MPS-based representation of Boolean
functions, MPS-based approximations to classical computational
problems have been attracting increasing attention from the
mathematics, computer science, and, more recently, physics communities
for more than a decade. In the classical context relevant to this
paper MPS representations were first proposed under the name {\it
  Tensor Train-} or TT-representations in a series of papers by
Oseledets and collaborators \cite{Oseledets2011, Dolgov2014,
  Lubich2015, Cichocki2017, KhrulkovHO19, novikov2022,
  sengupta2022}. The principal message of those papers is that
high-dimensional matrices appearing in the formulation of certain
numerical problems, inaccessible by traditional techniques, can be
compressed via a MPS representation, resulting in a (sometimes
exponential) speedup of the solution to the initial problem.

Since the original proposal, quantum-inspired MPS/TT approximations
have been applied, among others, to high-precision solutions to: (i)
nonlinear partial differential equations \cite{Khoromskij2015}
including the Vlasov \cite{Kormann2015, Einkemmer2018,
  Allmann-Rahn2022, ye2024}, Navier-Stokes \cite{kornev2023,
  peddinti2023, Kiffner2023}, and nonlinear Schr{\"o}dinger
\cite{Lubasch2018} equations; (ii) multigrid methods
\cite{Lubasch2018, Bolten2018}; (iii) high-resolution representations
of univariate and multivariate \cite{Tichavsky2024} functions; (iv)
high-dimensional integration \cite{Vysotsky2021}, convolutions
\cite{Su2020}, and Fourier transforms \cite{Wahls2014}; and (v)
interpolation and extrapolation of functions
\cite{Ritter2024}. Especially noteworthy are the applications to
machine learning algorithms \cite{Stoudenmire2016, Novikov2018,
  Sozykin2022}, an area in which the number of papers and the breadth
of problems explored via MPS-based methods are continuing to grow
\cite{fernandez2024}. The successes of MPS-based classical
computations hinge on the fact that the maximum matrix ranks (bond
dimensions) generated by the MPS representations of these classical
problems -- a measure of entanglement in the quantum context -- remain
small or moderate.

The contribution of this paper was motivated by our recently proposed
approach to classical computation on encrypted data we refer to as
Encrypted Operator Computing (EOC) \cite{chamon2022}. Within the EOC
scheme operations on encrypted data are carried out via an encrypted
circuit based on reversible computation \cite{Fredkin1982}. The
encrypted circuit is represented as a concatenation of a polynomial
number of ``chips'', $n$-input/$n$-output reversible functions, the
outputs of which are expressed as polynomial-sized (ordered) Binary
Decision Diagrams (BDDs). For a given variable order, since BDDs are
normal forms, they only expose the functionality of the chip but hide
its original circuit implementation. BDDs were invented by Bryant in
1986 \cite{Bryant1986} and were referred to by Knuth in his 2008
lectures as ``one of the only really fundamental data structures that
came out in the last twenty-five years'' \cite{youtube}. The
motivation for developing the BMP representation was the barrier we
encountered in using and modifying CUDD~\cite{somenzi1998cudd}, a
complex, highly optimized library of tools for manipulating BDDs, that
is no longer being maintained and the inner workings of which are
difficult to access by outsiders to the area of BDDs. As will be
described in some details in the body of this paper, BMPs only involve
simple linear algebra and are more flexible and easier to implement
than BDDs. The translation from BMPs to BDDs will also provide
insights into the special features and limitations of each of these
data structures.

It is important to stress that, unlike MPS representations of quantum
states, which are usually used as approximations of the full many-body
wave function, the BMP representation is in principle exact, but can
be efficiently implementable if and only if the maximum bond dimension
remains polynomial in the number of variables. As in the quantum case,
determining what features of algorithms or functions control the
behavior of bond dimension in the course of implementation remains an
unsolved problem. What is clear is that generically, the maximum bond
dimension of BMPs is very sensitive to the order of Boolean variables
and that finding the optimal variable ordering is a NP-complete problem, a result we take over from the BDD
literature~\cite{BolligWegener1996}.

This paper is organized as follows. In Sec.~\ref{sec:bmp}, we provide
a detailed definition of BMPs and in Sec.~\ref{sec:normal-form} we
show that, when fully compressed, they represent normal forms of
Boolean functions. In Sec.~\ref{sec:operations} we present the basic
operations needed in the synthesis and manipulation of BMPs; and in
Sec.~\ref{sec:BDD} we discuss the translation between BMPs and
BDDs. In Sec.~\ref{sec:complexity}, we discuss the computational
complexity of BMPs reflected in the maximum bond dimension associated
with different orderings of Boolean variables and compare the two
proposed implementations (via direct products and direct sums,
respectively) for the APPLY operation, arguably the most frequently
used operation on BMPs (see below). In Sec.~\ref{sec:conclusions} we
provide brief conclusions. Finally, some of the implementation details
of working with row-switching matrices and BMPs are given in the
Appendix.

%%%%%%%%%%%%%%%%%%%%%%%%%%%%%%%%%%%%%%%%%%%%%%%%%%%%%%%%%%%%%%%%%%%%%%%%%
\section{The Boolean Matrix Product (BMP) Representation}
\label{sec:bmp}

Consider a column vector $\bf F(\bf{x})$ of $m$ Boolean functions $f_i
(x_{n-1}, x_{n-2}, ...,x_{0}), i=1,...,m$, defined on $n$ Boolean
variables, ${\bf{x}} = (x_{n-1}, x_{n-2},...,x_1,..., x_0)$, where
$x_i\in \{0,1\}$ and $i=0,1,..,n-1$. The approach is based on an
iterative use of the Shannon decomposition of Boolean functions, one
Boolean variable at a time: we start with $x_{n-1}$ and expand each of
the $m$ Boolean functions $f_i$ as $f_i (x_{n-1}, x_{n-2}, ...,x_{0})
= \bar{x}_ {n-1} \,f_i (0, x_{n-2},...,x_{0}) + x_{n-1}\,
f_i(1,x_{n-2},...,x_{0})$, where $\bar{x}_i=1-x_i$ is the negation of
the variable $x_i$. Notice that, in the Shannon decomposition, we exploit that $f$'s and $x$'s are
binary and use integer multiplication ($\cdot$) and
addition ($+$) instead of AND and XOR.

The BMP construction then proceeds in three steps: first, the Shannon
decomposition with respect to $x_{n-1}$ is expressed in (column)
vector form, as:
\begin{equation}
\label{eq:shannon_decomp}
\begin{split}
{\bf{F}(\bf{x})}&\equiv
\left(\begin{array}{c}
    f_1 (x_{n-1}, x_{n-2}, \ldots, x_{1}, x_0)\\
    f_2 (x_{n-1}, x_{n-2}, \ldots, x_{1}, x_0)\\
    \vdots\\
    f_m (x_{n-1}, x_{n-2}, \ldots, x_{1}, x_0)
  \end{array}\right)\\
  &=
  \left(\begin{array}{cccccccccc}
  \bar{x}_{n-1} &0&0&\ldots&0&x_{n-1}&0&0&\ldots&0\\
  0&\bar{x}_{n-1}&0&\ldots&0&0&x_{n-1}&0&\ldots&0\\
  \vdots\\
  0&0&\ldots&\bar{x}_{n-1}&0&0&0&\ldots&x_{n-1}&0\\
  0&0&\ldots&0&\bar{x}_{n-1}&0&0&\ldots&0&x_{n-1}
  \end{array}\right)
  \left(\begin{array}{c}
    f_1 (0, x_{n-2}, \ldots, x_{0})\\
    f_2 (0, x_{n-2}, \ldots, x_{0})\\
    \vdots\\
    f_m (0, x_{n-2}, \ldots, x_{0})\\
    f_1 (1, x_{n-2}, \ldots, x_{0})\\
    f_2 (1, x_{n-2}, \ldots, x_{0})\\
    \vdots\\
    f_m (1, x_{n-2}, \ldots, x_{0})
  \end{array}\right)\\
 &\equiv \left[{\bf{S}} (x_{n-1})\right ] _{m\times 2m}
  \left[{\bf{F}}^{({n-1})}(x_{n-2}, \ldots, x_{0})\right]_{2m\times 1}.
\end{split}
\end{equation}
Here ${\bf{S}} (x)$, which we refer to as the ``Shannon'' matrix, can
be expressed in compact form as the Kronecker tensor product
\begin{equation}
\label{eq:shannon}
\left[{\bf{S}} (x)\right ]_{m\times 2m} = \left[ \bar{x} \; x
\right]_{1\times 2} \otimes \left[{\mathbb{1}}\right ]_{m\times m},
\end{equation}
where ${\mathbb{1}}$ is the identity matrix.

The second step is to remove the redundancy in
${\bf{F}}^{({n-1})}(x_{n-2}, \ldots, x_{0})$ that arises when
pairs of functions happen to be equal, i.e.,
$f_k (x_{n-1} = v, x_{n-2},...,x_{0}) \equiv f_l(x_{n-1} =
w,x_{n-2},...,x_{0})$ for any $k,l \in \{1,\dotsc,m\}$ and
$v,w \in \{0,1\}$. Such repeated rows can be eliminated by introducing
a $2m\times m_{n-1}$ ``row switching" (RS) matrix
${\bf{U}}^{({n-1})}$, which has a single 1 in every one of its $2m$
rows, which picks out the appropriate function from the compressed set
of $m_{n-1}$ independent functions as follows:
\begin{equation}
\label{eq:compression}
\left[{{\bf{F}}^{({n-1})}(x_{n-2}, \ldots, x_{0})}\right]_{2m\times 1} =
\left[{\bf{U}}^{({n-1})}\right]_{2m\times m_{n-1}}
\left[{\tilde{\bf{F}}}^{({n-1})}(x_{n-2}, \ldots, x_{0})\right]_{m_{n-1}\times 1},
\end{equation}
where $m_{n-1}\le 2m$. The RS matrix ${\bf{U}}^{({n-1})}$ can be
split into two $m\times m_{n-1}$ blocks, ${\bf{M}}^{({n-1})} (0)$ and
${\bf{M}}^{({n-1})} (1)$,
\begin{equation}
\label{eq:RS}
{\left[{\bf{U}}^{({n-1})}\right]_{2m\times{m_{n-1}}}} \equiv
\left[ \begin{array}{c} {\bf{M}}^{({n-1})} (0)\\ {\bf{M}}^{({n-1})}
    (1)
\end{array} \right]_{2m\times{m_{n-1}}},
\end{equation}
which, using the Shannon matrix in Eq.~\eqref{eq:shannon}, define the Shannon decomposition of the matrix
,
\begin{equation}
\label{eq:M_matrix}
\left[{\bf{M}}^{({n-1})} (x_{n-1})\right]_{m\times{m_{n-1}}} = \left[
  {\bf{S}} (x_{n-1})\right]_{m\times 2m}
\left[{\bf{U}}^{({n-1})}\right]_{2m \times m_{n-1}}
\,.
\end{equation}
By using Eqs.~\eqref{eq:compression} and~\eqref{eq:M_matrix}, we
can express the Shannon decomposition Eq.~\eqref{eq:shannon_decomp} as
\begin{equation}
{\bf{F}(\bf{x})} = \left[{\bf{M}}^{({n-1})} (x_{n-1})\right]_{m\times m_{n-1}}
\left[{\tilde{\bf{F}}}^{({n-1})}(x_{n-2}, \ldots, x_{0})\right]_{m_{n-1}\times 1}
\;.
\end{equation}
As the last step in the construction, the Shannon decomposition and
compression via the elimination of duplicate rows are iterated through
the remaining $(n-1)$ variables, leading to the final BMP
representation in terms of RS matrices ${\bf{M}}^{(i)} (x_i)$,
\begin{equation}
{{\bf{F}}(\bf{x})} = \left[{\bf{M}}^{({n-1})}
  (x_{n-1})\right]_{m\times m_{n-1}} \left[{\bf{M}}^{({n-2})}
  (x_{n-2})\right]_{m_{n-1}\times m_{n-2}}\cdots \left[{\bf{M}}^{(1)}
  (x_{1})\right]_{m_{2}\times m_{1}} \left[{\bf{M}}^{(0)}
  (x_{0})\right]_{m_{1}\times 2}\;\left[ {\bf R} \right]_{2\times 1},
\label{eq:bmp}
\end{equation}
with the terminal vector,
\begin{equation}
\label{eq:R}
{\bf R}=\left(\begin{array}{c}
    0\\
    1\\
  \end{array}\right)
  \end{equation}
representing the two possible constant Boolean functions at the end of
the matrix train.

The rank $m_i$ of the RS matrix $\left[{\bf{M}}^{(i)}
  (x_{i})\right]_{m_{i+1}\times m_{i}}$ -- also referred to as the
``bond dimension" (associated with the ``link" between matrices $i$
and $i-1$ along the matrix train) -- satisfies the inequality $m_i \le
2 m_{i+1}$.  This implies that the upper bound on the bond
dimension at the $i$th site across the BMP scales exponentially,
$d_i\le 2^{i+1}$, underscoring the fact that the BMP is only useful in
representing Boolean functions for which the bond dimension, $d_i$,
can be kept under control, i.e., when $d_i$ scales polynomially with
$i$ or when the number of Boolean variables is sufficiently small.

%%%%%%%%%%%%%%%%%%%%%%%%%%%%%%%%%%%%%%%%%%%%%%%%%%%%%%%%%%%%%%%%%%%%%%%%%

\subsection{An example}
\label{sec:example}
%%%%%%%%%%%%%%%%%%%%%%%%%%%%%%%%%%
% Step 1
%%%%%%%%%%%%%%%%%%%%%%%%%%%%%%%%%%

Consider the Boolean function $h({\bf{x}}) = \bar x_2 x_1 x_0$, which
we use as a concrete example of the BMP construction described above. In this
example with $n=3$ and $m=1$, we write
\begin{align}
  {\bf F}({\bf x}) = (h(x_2,x_1,x_0))
  =\left[{\bf{S}} (x_{2})\right ] _{1\times 2}
  \left[{\bf{F}}^{({2})}(x_{1}, x_{0})\right]_{2\times 1}
  \;,
\end{align}
with
\begin{align}
  \left[{{\bf{F}}^{({2})}(x_1, x_0)}\right]_{2\times 1} =
  \left(\begin{array}{c}
    h (0, x_{1}, x_{0})\\
    h (1, x_{1}, x_{0})
  \end{array}\right)
  =
  \left(\begin{array}{c}
    x_{1}\,x_{0}\\
    0
  \end{array}\right)
  =
  \left(\begin{array}{cc}
    1&0 \\ \hline
    0&1
  \end{array}\right)
  \;
  \left(\begin{array}{c}
    x_{1}\,x_{0}\\
    0
  \end{array}\right)
  =
  \left[{\bf{U}}^{({2})}\right]_{2\times 2}
  \left[{\tilde{\bf{F}}}^{({2})}(x_{1}, x_{0})\right]_{2\times 1}
  \;.
\end{align}
From the upper and lower blocks of
$\left[{\bf{U}}^{({2})}\right]_{2\times 2}$, which we separate by a
horizontal line in the above equation, we extract
\begin{align}
  \left[{\bf{M}}^{({2})} (0)\right]_{1\times{2}}
  =
  \left(\begin{array}{cc}
    1&0
  \end{array}\right)  
  \quad
  \text{and}
  \quad
  \left[{\bf{M}}^{({2})} (1)\right]_{1\times{2}}
  =
  \left(\begin{array}{cc}
    0&1
  \end{array}\right)
  \;,
\end{align}
or equivalently,
\begin{align}
  \left[{\bf{M}}^{({2})} (x_{2})\right]_{1\times{2}}
  =
  \left(\begin{array}{cc}
    \bar x_{2}&x_{2}
  \end{array}\right)
  \;.
\end{align}
%%%%%%%%%%%%%%%%%%%%%%%%%%%%%%%%%%
% Step 2
%%%%%%%%%%%%%%%%%%%%%%%%%%%%%%%%%%

We then proceed with the decomposition of 
\begin{align}
  \left[{\tilde{\bf{F}}}^{({2})}(x_{1}, x_{0})\right]_{2\times 1}
  =
  \left(\begin{array}{c}
    x_{1}\,x_{0}\\
    0
  \end{array}\right)
  =\left[{\bf{S}} (x_{1})\right ] _{2\times 4}
  \left[{\bf{F}}^{({1})}(x_{0})\right]_{4\times 1}
  \;,
\end{align}
with
\begin{align}
  \left[{{\bf{F}}^{({1})}(x_0)}\right]_{4\times 1}
  =
  \left(\begin{array}{c}
    0\cdot x_{0}\\
    0 \\
    1\cdot x_{0}\\
    0
  \end{array}\right)
  =
  \left(\begin{array}{c}
    0 \\
    0 \\
    x_{0}\\
    0
  \end{array}\right)
  =
  \left(\begin{array}{cc}
    1&0\\
    1&0\\ \hline
    0&1\\
    1&0
  \end{array}\right)
  \;
  \left(\begin{array}{c}
    0\\
    x_{0}
  \end{array}\right)
  =
  \left[{\bf{U}}^{({1})}\right]_{4\times 2}
  \left[{\tilde{\bf{F}}}^{({1})}(x_{0})\right]_{2\times 1}
  \;,
\end{align}
from which we read the upper and lower blocks of 
$\left[{\bf{U}}^{({1})}\right]_{4\times 2}$ (again separated by a
horizontal line in the above equation),
\begin{align}
  \left[{\bf{M}}^{({1})} (0)\right]_{1\times{2}}
  =
  \left(\begin{array}{cc}
    1&0\\
    1&0
  \end{array}\right)  
  \quad
  \text{and}
  \quad
  \left[{\bf{M}}^{({1})} (1)\right]_{1\times{2}}
  =
  \left(\begin{array}{cc}
    0&1\\
    1&0
  \end{array}\right)
  \;,
\end{align}
or equivalently,
\begin{align}
  \left[{\bf{M}}^{({1})} (x_{2})\right]_{1\times{2}}
  =
  \left(\begin{array}{cc}
    \bar x_{1}&x_{1}\\
    1&0
  \end{array}\right)
  \;.
\end{align}
%%%%%%%%%%%%%%%%%%%%%%%%%%%%%%%%%%
% Step 3
%%%%%%%%%%%%%%%%%%%%%%%%%%%%%%%%%%

Finally, we decompose
\begin{align}
  \left[{\tilde{\bf{F}}}^{({1})}(x_{0})\right]_{2\times 1}
  =
    \left(\begin{array}{c}
    0\\
    x_{0}
  \end{array}\right)
  =\left[{\bf{S}} (x_{0})\right ] _{2\times 4}
  \left[{\bf{F}}^{({0})}\right]_{4\times 1}
  \;,
\end{align}
with
\begin{align}
  \left[{{\bf{F}}^{({1})}(x_0)}\right]_{4\times 1}
  =
  \left(\begin{array}{c}
    0 \\
    0 \\
    0 \\
    1
  \end{array}\right)
  =
  \left(\begin{array}{cc}
    1&0\\
    1&0\\ \hline
    1&0\\
    0&1
  \end{array}\right)
  \;
  \left(\begin{array}{c}
    0\\
    1
  \end{array}\right)
  =
  \left[{\bf{U}}^{({0})}\right]_{4\times 2}
  \left[{\bf{R}}\right]_{2\times 1}
  \;,
\end{align}
yielding the last row-switching matrices
\begin{align}
  \left[{\bf{M}}^{({0})} (0)\right]_{1\times{2}}
  =
  \left(\begin{array}{cc}
    1&0\\
    1&0
  \end{array}\right)  
  \quad
  \text{and}
  \quad
  \left[{\bf{M}}^{({0})} (1)\right]_{1\times{2}}
  =
  \left(\begin{array}{cc}
    1&0\\
    0&1
  \end{array}\right)
  \;,
\end{align}
or
\begin{align}
  \left[{\bf{M}}^{({0})} (x_{0})\right]_{1\times{2}}
  =
  \left(\begin{array}{cc}
    1&0\\
    \bar x_{0}&x_{0}
  \end{array}\right)
  \;.
\end{align}

%%%%%%%%%%%%%%%%%%%%%%%%%%%%%%%%%%
% Summary
%%%%%%%%%%%%%%%%%%%%%%%%%%%%%%%%%%

Finally, we assemble the BMP for our simple but concrete example, 
the Boolean function $h({\bf{x}}) = \bar x_2 x_1 x_0$:
\begin{equation}
    h(x_2,x_1,x_0) =
    \begin{pmatrix}
        \overline{x}_2 & x_2
    \end{pmatrix}
    \begin{pmatrix}
        \overline{x}_1 & x_1 \\ 1 & 0
    \end{pmatrix}
    \begin{pmatrix}
        1 & 0 \\ \overline{x}_0 & x_0
    \end{pmatrix}
    \begin{pmatrix}
        0 \\ 1
    \end{pmatrix}.
\end{equation}
%
%%%%%%%%%%%%%%%%%%%%%%%%%%%%%%%%%%%%%%%%%%%%%%%%%%%%%%%%%%%%%%%%%%%%%%%%%

%%%%%%%%%%%%%%%%%%%%%%%%%%%%%%%%%%%%%%%%%%%%%%%%%%%%%%%%%%%%%%%%%%%%%%%%%
\section{Compressed BMPs are canonical normal forms}
\label{sec:normal-form}

The BMP in Eq.~(\ref{eq:bmp}), built via the process outlined above,
is a canonical normal form, i.e., for a given order of the variables,
$(x_{n-1},\dots,x_1,x_0)$, Eq.~(\ref{eq:bmp}) provides a unique
compressed matrix product representation of a vector of Boolean
functions, up to permutations of lines and columns of the ${\bf{M}}$
matrices. This property follows from two elements of the derivation of
the BMP form in Eq.~(\ref{eq:shannon_decomp}): (a) the construction
proceeds iteratively, one variable at the time; and (b) once a
variable order is chosen, at every stage of the iteration both the
Shannon decomposition and the compression via elimination of dependent
rows are unique, up to a permutation of the unique rows.

The redundancy arising from permutations of the rows and columns of
${\bf M}$ constitutes, in the language of physics, a form of gauge
symmetry. The BMP representation -- expressed as a product of
row-switching matrices -- is preserved under the insertion of a
permutation matrix and its inverse, ${{\bf P}}^{-1} \;{{\bf P}} =
{\bf 1}$, between consecutive ${\bf M}$ matrices, namely:
\begin{align}
{\bf{F}}({\bf {x}}) &=
{{\bf{M}}}^{({n-1})}(x_{n-1})\;
{{\bf{M}}}^{({n-2})} (x_{n-2}) \cdots
%{{\bf{M}}}^{({2})} (x_{2}) \; {{\bf{M}}}^{({1})} (x_{1}) \;
{{\bf{M}}}^{({0})} (x_{0}) \; {\bf R}
\nonumber\\
&=
\left[{{\bf{M}}}^{({n-1})}(x_{n-1})\;{{\bf P}^{-1}_{n-1}}\right] \;
\left[{{\bf P}^{\,}_{n-1}}\;{{\bf{M}}}^{({n-2})} (x_{n-2})\;{{\bf P}^{-1}_{n-2}}\right]
\cdots
%{{\bf{M}}}^{({2})} (x_{2}) \; {{\bf{M}}}^{({1})} (x_{1}) \;
\left[{{\bf P}^{\,}_{1}}\; {{\bf{M}}}^{({0})} (x_{0})\right] \; {\bf R}
\label{eq:perm_bmp}\\
&=
{\tilde{\bf{M}}}^{({n-1})}(x_{n-1}) \;
{\tilde{\bf{M}}}^{({n-2})} (x_{n-2}) \cdots
%{{\bf{M}}}^{({2})} (x_{2}) \; {{\bf{M}}}^{({1})} (x_{1}) \;
{\tilde{\bf{M}}}^{({0})} (x_{0}) \; {\bf R}
\nonumber
\end{align}
where the transformations ${\tilde{\bf{M}}}^{({k})} (x_{k}) = {{\bf
    P}^{\,} _{k+1}}\; {\bf{M}}^{({k})} (x_{k}) \; {{\bf P}^{-1}_k}$,
involving permutations of rows and columns of the row switching
matrices ${\bf{M}}^{({k})} (x_{k})$, preserve the row switching
property and do not change the rank of the original matrices. The
redundancy can be removed by choosing any ``lexicographic'' ordering of
the unique rows of ${\tilde{\bf{F}}}$ in the decomposition in
Eq.~\eqref{eq:compression}, making the BMP decomposition of the Boolean vector ${\bf{F}}(\bf {x})$
unique.

%%%%%%%%%%%%%%%%%%%%%%%%%%%%%%%%%%%%%%%%%%%%%%%%%%%%%%%%%%%%%%%%%%%%%%%%%
\section{BMP Operations}
\label{sec:operations}

Here we present (binary) matrix product operations that enable
synthesis of Boolean circuits and general manipulations of Boolean
functions in the BMP representation. Before delving into the details
of BMP operations, we present simple Shannon decomposition
formulas that will be used below without further explanation. In
particular, the Shannon matrix of Eq.~\eqref{eq:shannon} and its
transpose,
\begin{equation}
\left[{{\bf{S}}^\top} (x)\right ] = \left[ \bar{x} \; x \right]^\top \otimes
\left[{\bf {1}}\right ],
\end{equation}
are used, respectively, in the column-by-column and row-by-row
decompositions of a Boolean matrix $\bf{M}({\bf{x}})$:
\begin{equation}
\label{eq:shannon_decompositions}
\left[{\bf{M}}({x})\right]_{n\times m} = \left[{\bf{S}} (x)
  \right]_{n\times 2n}\left[ \begin{array}{c}
    {\bf{M}}(0)\\ {\bf{M}}(1) \end{array} \right]_{2n\times m} =
\left[{\bf{M}}(0) \;\;\; {\bf{M}}(1) \right]_{n\times 2n} \left[
  {{\bf{S}}^\top} (x) \right]_{2n\times m}.
\end{equation}
%

%%%%%%%%%%%%%%%%%%%%%%%%%%%%%%%%%%%%%%%%%%%%%%%%%%%%%%%%%%%%%%%%%%%%%%%%%
\subsection{CLEAN}
\label{sec:clean}

Operating on or combining canonical form (compressed) BMPs results in
matrix products that are not generally in canonical form. Importantly,
manipulations of BMPs that can be performed efficiently (i.e., with
polynomial resources) lead to matrix products that can be compressed
into canonical form with a polynomial number of steps, an operation we
refer to as CLEAN.

Starting from a general BMP of a $p$ component vector Boolean
function, for which the matrices $\left[{\bf Q}^{(i)} (x_i)\right]$
are not necessarily row-switching,
\begin{equation}
\label{eq:bmp_gen}
\left[{\bf{G}}\right]_{p\times 1} =\left[{\bf{Q}}^{({n-1})}
  (x_{n-1})\right]_{p\times p_{n-1}}
\left[{\bf{Q}}^{({n-2})}(x_{n-2})\right]_{p_{n-1}\times p_{n-2}}\ldots
\left[{\bf{Q}}^{(1)} (x_{1})\right]_{p_2 \times p_1}
\left[{\bf{Q}}^{(0)} (x_{0})\right]_{p_1\times p_0} \left[{\bf T}
  \right]_{p_0 \times 1},
\end{equation}
the CLEAN operation involves sequentially compressing each of the
matrices one-at-a-time traversing the matrix train from left-to-right
(LTR) and then from right-to-left (RTL). Here we present explicit
constructions of the LTR- and RTL-cleaning procedures based on the two
steps we followed in deriving the BMP representation of
Eq.~(\ref{eq:bmp}), namely a Shannon decomposition of the matrix
under consideration followed by the elimination of redundant rows
(compression).

%%%%%%%%%%%%%%%%%%%%%%%%%%%%%%%%%%%%%%%%%%%%%%%%%%%%%%%%%%%%%%%%%%%%%%%%%
\subsubsection{LTR cleaning}
\label{sec:LTR-clean}

We proceed by considering the Shannon decomposition of the left-most
matrix in the train, ${\bf{Q}}^{({n-1})} (x_{n-1})$,
\begin{equation}
\label{eq:LTR-cleaning}
\left[{\bf{Q}}^{({n-1})} (x_{n-1})\right]_{p\times p_{n-1}}
=\left[{\bf{S}}(x_{n-1}) \right]_{p\times 2p}\; \left[\begin{array}{c}
    {\bf{Q}}^{({n-1})} (0)\\ {\bf{Q}}^{({n-1})} (1)
\end{array}\right]_{2p\times p_{n-1}}.
\end{equation}
We note that Eq.~\eqref{eq:LTR-cleaning} describes the
column-by-column Shannon decomposition of the matrix
${\bf{Q}}^{({n-1})} (x_{n-1})$ which one could also regard as the
inner product of two auxiliary vectors: $(\bar{x}, x)$ and
$({\bf{Q}}^{({n-1})} (0), {\bf{Q}}^{({n-1})} (1))$. For consistency we
prefer to use the decomposition used in deriving Eq.~\eqref{eq:bmp} or
its row-by-row (transpose) version in which $\bar{x}$ and $x$ are
simple Boolean variables.

We next implement the compression step by using a RS matrix
${\tilde{\bf{U}}}^{({n-1})}$ with $\tilde{p}_{n-1}\le 2p$ to write
\begin{equation}
\label{eq:LTR_compression}
 \left[\begin{array}{c} {\bf{Q}}^{({n-1})} (0)\\ {\bf{Q}}^{({n-1})}
     (1) \end{array} \right]_{2p\times p_{n-1}} = \left[
   {\tilde{\bf{U}}}^{({n-1})} \right]_{2p\times \tilde{p}_{n-1}}
 \left[ \tilde{\bf{Q}}^{({n-1})} \right]_{\tilde{p}_{n-1} \times
   p_{n-1}}.
\end{equation}
As in Eq.~(\ref{eq:M_matrix}), we split the $2p\times
\tilde{p}_{n-1}$ RS matrix $\left[{\tilde{\bf{U}}}^{({n-1})}\right]$
into two $p\times \tilde{p} _{n-1}$ blocks,
\begin{equation}
\label{eq:tildeM_matrix}
\left[\tilde{{\bf{U}}}^{({n-1})}\right]_{2p\times \tilde{p}_{n-1}}
\equiv \left[\begin{array}{c} {\tilde{\bf{M}}}^{({n-1})}
    (0)\\ {\tilde{\bf{M}}}^{({n-1})} (1)
\end{array}\right]_{2p\times \tilde{p}_{n-1}},
\end{equation}
which, together with Eqs.~\eqref{eq:LTR-cleaning} and
\eqref{eq:LTR_compression}, allows us to write
\begin{eqnarray}
\left[ {\bf{Q}}^{({n-1})} (x_{n-1}) \right]_{p\times p_{n-1}} & = &
\left[ {\bf{S}}(x_{n-1}) \right]_{p\times 2p} \left[ \begin{array}{c}
    {\tilde{\bf{M}}}^{({n-1})} (0) \\ {\tilde{\bf{M}}}^{({n-1})}
    (1) \end{array} \right]_{2p\times \tilde{p}_{n-1}} \left[
  \tilde{\bf{Q}}^{({n-1})} \right]_{\tilde{p}_{n-1}\times p_{n-1}}
\nonumber \\ & = & \left[ \tilde{\bf{M}}^{(n-1)} (x_{n-1})
  \right]_{p\times \tilde{p}_{n-1}} \left[ \tilde{\bf{Q}}^ {({n-1})
  }\right]_{\tilde{p}_{n-1} \times p_{n-1}}.
\label{eq:QvsM}
\end{eqnarray}
Absorbing the matrix $\tilde{\bf{Q}}^ {({n-1})}$ to the right we can
now proceed with the LTR-cleaning process of matrix
\begin{equation*}
{\bf{Q'}}^ {({n-2})} (x_{n-2}) = \tilde{\bf{Q}}^ {({n-1})} \;{\bf{Q}}^
{({n-2})} (x_{n-2}),
\end{equation*}
a process we continue iterating to the right along the matrix train of
Eq.~\eqref{eq:bmp_gen}. Finally, the last $\tilde{\bf{Q}}^{(0)}$
matrix emerging from applying the cleaning process to ${\bf{Q}}^
{({0})} (x_{0})$ is absorbed into the terminal vector
\begin{equation*}
\left[{\bf
    T'}\right] _{\tilde{p}_0\times 1}=
\left[\tilde{\bf{Q}}^{(0)}\right]_{\tilde{p}_0\times p_0}\left[{\bf
    T}\right] _{{p_0}\times 1},
\end{equation*}
resulting in the BMP decomposition
\begin{equation}
\left[{\bf{G}}\right]_{p\times 1} =\left[\tilde{\bf{M}}^{({n-1})}
  (x_{n-1})\right]_{p\times \tilde{p}_{n-1}}
\left[\tilde{\bf{M}}^{({n-2})}(x_{n-2})\right]_{\tilde{p}_{n-1}\times
  \tilde{p}_{n-2}}\ldots \left[\tilde{\bf{M}}^{(1)}
  (x_{1})\right]_{\tilde{p}_2 \times \tilde{p}_1}
\left[\tilde{\bf{M}}^{(0)} (x_{0})\right]_{\tilde{p}_1\times
  \tilde{p}_0} \left[{\bf T}^\prime \right]_{\tilde{p}_0 \times 1}.
\end{equation}
%

%%%%%%%%%%%%%%%%%%%%%%%%%%%%%%%%%%%%%%%%%%%%%%%%%%%%%%%%%%%%%%%%%%%%%%%%%
\subsubsection{RTL cleaning}
\label{sec:RTL-clean}

The first step in RTL-cleaning is the compression of the terminal
vector ${\bf{T'}}$, accomplished through the use of a
$\tilde{p}_0\times 2$ RS matrix ${{\tilde{\tilde{\bf{U}}}}_T}$ via
${\bf{T'}} = {{\tilde{\tilde{\bf{U}}}}_T} \;\bf{R}$, with
${\bf{R}}$ given by Eq.~(\ref{eq:R}). Once we absorb
${{\tilde{\tilde{\bf{U}}}}_T}$ into the first matrix of the matrix
train to the left via $ {\bf{Q'}}^{(0)} (x_{0}) = \tilde{\bf{M}}^{(0)}
(x_{0}) \;{{\tilde{\tilde{\bf{U}}}}_T}$ we proceed with
implementing the RTL-cleaning process one-variable-at-a-time starting
with ${\bf{Q'}}^{(0)} (x_{0})$.
%%(Note that, had a LTR-cleaning been already applied to the BMP of
%%Eq.~\eqref{eq:bmp_gen} the first matrix on the left of the terminal
%%node would be ${\tilde{\bf{M}}}^{(0)} (x_0)$. Here we assume the
%%LTR-cleaning had not yet been applied.)
The first step in the RTL-cleaning process, the Shannon decomposition
of the matrix ${\bf{Q'}}^{(0)} (x_{0})$, is now implemented
row-by-row:
\begin{equation}
\label{eq:shannon2}
\left[{\bf{Q'}}^{(0)} (x_{0})\right]_{\tilde{p}_1\times \tilde{p}_0} =
\left[{\bf{Q'}}^{(0)} (0) \; \; {\bf{Q'}}^{(0)}
  (1)\right]_{\tilde{p}_1\times 2 \tilde{p}_0} \left[{{\bf{S}}^\top}
  (x_0)\right ]_{2 \tilde{p}_0\times \tilde{p}_0}.
\end{equation}
As in the LTR-cleaning process, the second step is the removal of
redundant rows (compression) of the matrix $\left[{\bf{Q'}}^{(0)} (0)
  \; \; {\bf{Q'}}^{(0)} (1)\right]_{\tilde{p}_1\times 2 \tilde{p}_0} $
by using an RS matrix $\tilde{\tilde{\bf{U}}}^{(0)}$:
\begin{equation}
\label{eq:RTL_compression}
\left[{\bf{Q'}}^{(0)} (0) \; {\bf{Q'}}^{(0)}
  (1)\right]_{\tilde{p}_1\times 2 \tilde{p}_0} =
\left[{\tilde{\tilde{\bf{U}}}}^{(0)}\right]_{\tilde{p}_1 \times
  \tilde{\tilde{p}} _1} \left[{\tilde{\tilde{\bf{M}}}}^{(0)} (0) \;\;
  {\tilde{\tilde{\bf{M}}}}^{(0)} (1)\right]_{\tilde{\tilde{p}}_1
  \times 2 \tilde{p}_0},
\end{equation}
with $\tilde{\tilde{p}}_1 \le
\tilde{p}_1$. Equations.~\eqref{eq:shannon2} and
\eqref{eq:RTL_compression} can now be used together with the
column-by-column Shannon decomposition to write
\begin{equation}
\label{RTL_cleaning}
\left[ {\bf{Q'}}^{(0)} (x_{0})\right]_{\tilde{p}_1\times \tilde{p}_0}
= \left[ {\tilde{\tilde{\bf{U}}}}^{(0)} \right]_{\tilde{p}_1 \times
  \tilde{\tilde{p}}_1} \left[ {\tilde{\tilde{\bf{M}}}}^{(0)} (0) \; \;
  {\tilde{\tilde{\bf{M}}}}^{(0)} (1) \right]_{\tilde{\tilde{p}}_1
  \times 2\tilde{p}_0} \left[ {{\bf{S}}^\top} (x_0)\right]_{2\tilde{p}_0
  \times \tilde{p}_0} =
\left[{\tilde{\tilde{\bf{U}}}}^{(0)}\right]_{\tilde{p}_1 \times
  \tilde{\tilde{p}}_1} \left[{\tilde{\tilde{\bf{M}}}}^{(0)}
  (x_0)\right]_{\tilde{\tilde{p}}_1 \times \tilde{p}_0} .
\end{equation}
Absorbing ${\tilde{\tilde{\bf{U}}}}^{(0)}$ into the next matrix on the
left, $\tilde{\bf{M}}^{(1)}(x_1) \rightarrow {\bf{Q'}}^{(1)}(x_1) =
\tilde{\bf{M}}^{(1)} (x_1) \;{\tilde{\tilde{\bf{U}}}}^{(0)}$, we can
now proceed right-to-left iteratively and repeat the RTL-cleaning
process starting with ${\bf{Q'}}^{(1)} ({x_1})$. Note that a single
round of RTL-cleaning leaves the left-most matrix
uncompressed. Moreover, one round of RTL-cleaning on an arbitrary BMP
does not guarantee that the compressed matrices are row
switching. Reaching a canonical form requires at least one round of
LTR-cleaning.

%\vspace{.3cm}
%\noindent
%{\bf 2. APPLY}
%\vspace{.3cm}

%%%%%%%%%%%%%%%%%%%%%%%%%%%%%%%%%%%%%%%%%%%%%%%%%%%%%%%%%%%%%%%%%%%%%%%%%
\subsection{APPLY}
\label{sec:apply}

Given the canonical BMP representations of two Boolean functions,
$f({\bf{x}})\equiv f(x_{n-1}, x_{n-2}, ...., x_2, x_1, x_0)$ and
$g({\bf{x}})\equiv g(x_{n-1}, x_{n-2}, ...., x_2, x_1, x_0)$, the
APPLY operation constructs the BMP representation of a Boolean
function $h(f,g)$. There are two ways to implement APPLY: one based on
direct products and another based on direct sums.

%%%%%%%%%%%%%%%%%%%%%%%%%%%%%%%%%%%%%%%%%%%%%%%%%%%%%%%%%%%%%%%%%%%%%%%%%
\subsubsection{Direct product}
\label{sec:direct-prod}

Starting with the Shannon expansion of the function $h(f,g)$, we write
\begin{equation}
\label{hfg}
\begin{split}
h(f,g) & = \overline{f({\bf x})} \; \overline{g({\bf x})} \; h(0,0) +
\overline{f({\bf x})} \; g({\bf x}) \; h(0,1) + f({\bf x}) \;
\overline{g({\bf x})} \; h(1,0) + f({\bf x}) \; g({\bf x}) \; h(1,1)
\\ & = \left[ \left( \overline{f({\bf x})} \; \; f({\bf x}) \right)
  \otimes \left( \overline{g({\bf x})} \; \; g({\bf x}) \right)
  \right] \; \left(\begin{array}c h(00)\\ h(01)\\ h(10)\\ h(11)
\end{array}
\right)
\end{split}
\end{equation}
Now, using the BMP representations of $f$ and $g$,
\begin{equation}
\label{eq:fg-bmp}
\begin{split}
f({\bf x}) & = {\bf F}^{(n-1)} (x_{n-1}) \;{\bf F}^{(n-2)}
(x_{n-2}) \cdots {\bf F}^{(1)} (x_{1}) \;{\bf F}^{(0)} (x_{0})
\;{\bf R} \\ g({\bf x}) & = {\bf G}^{(n-1)} (x_{n-1}) \;{\bf
  G}^{(n-2)} (x_{n-2}) \cdots {\bf G}^{(1)} (x_{1}) \;{\bf
  G}^{(0)} (x_{0}) \;{\bf R},\\
\end{split}
\end{equation}
we can write
\begin{equation}
\begin{split}
\left( \overline{f(\bf{x})} \; \; f(\bf{x}) \right) & = {\bf
  F}^{(n-1)} (x_{n-1}) \;{\bf F}^{(n-2)} (x_{n-2}) \cdots {\bf
  F}^{(1)} (x_{1}) \;{\bf F}^{(0)} (x_{0}) \\ \left(
  \overline{g(\bf{x})} \; \; g(\bf{x}) \right) & = {\bf G}^{(n-1)}
(x_{n-1}) \;{\bf G}^{(n-2)} (x_{n-2}) \cdots {\bf G}^{(1)} (x_{1})
\;{\bf G}^{(0)} (x_{0}),\\
\end{split}
\end{equation}
and thus,
\begin{equation}
\label{eq:apply}
h(f,g) = {\bf H}^{(n-1)} (x_{n-1}) \;{\bf H}^{(n-2)} (x_{n-2})
\cdots {\bf H}^{(1)} (x_{1}) \;{\bf H}^{(0)} (x_{0}) \;
\left( \begin{array}{c} h(00)\\ h(01)\\ h(10)\\ h(11)
\end{array}
\right),
\end{equation}
where ${\bf H}^{(i)} (x_i) = {\bf F}^{(i)} (x_i) \otimes {\bf G}^{(i)}
(x_i)$. The canonical form for $h(f,g)$ is then obtained by applying
the CLEAN operation to the resulting BMP of Eq.~(\ref{eq:apply}).

%%%%%%%%%%%%%%%%%%%%%%%%%%%%%%%%%%%%%%%%%%%%%%%%%%%%%%%%%%%%%%%%%%%%%%%%%
\subsubsection{Direct sum}
\label{sec:direct-sum}

While the method based of direct products produces a compressed BMP
after the application of the CLEAN operation, the intermediate
matrices it generates can be prohibitively large. Here we present an alternative
and often more efficient method based on direct sum of the matrices in the BMP
representations of $f$ and $g$. This involves transforming the product
\begin{equation}
  \label{eq:direct-sum-apply-init}
    \begin{pmatrix}
        1 & 1
    \end{pmatrix} \;
    \left[ {\bf F}^{(n-1)}(x_{n-1}) \oplus {\bf G}^{(n-1)}(x_{n-1}) \right] \;
    \left[ {\bf F}^{(n-2)}(x_{n-2}) \oplus {\bf G}^{(n-2)}(x_{n-2}) \right]
    \cdots
    \left[ {\bf F}^{(0)}(x_{0}) \oplus {\bf G}^{(0)}(x_{0}) \right]
\end{equation}
into
\begin{equation}\label{eq:direct-sum-apply-final1}
    {\bf H}^{(n-1)}(x_{n-1}) \;{\bf H}^{(n-2)}(x_{n-2}) \;{\bf
      H}^{(1)}(x_{1}) \;{\bf H}^{(0)}(x_{0}) \;{\bf U}^{(0)}
\end{equation}
in a process similar to the LTR sweep of the CLEAN operation. Here,
the matrices ${\bf H}^{(i)}$ are row-switching, while the rows of
${\bf U}^{(0)}$ contain all the possible outputs of the product. Since
the initial product evaluates to
\begin{equation}
    \left( \overline{f({\bf x})} \; \; f({\bf x}) \; \;
    \overline{g({\bf x})} \; \; g({\bf x}) \right),
\end{equation}
each of these distinct outputs corresponds to a particular combination
of the values of $f({\bf x})$ and $g({\bf x})$. Therefore, replacing
the rows of ${\bf U}^{(0)}$ with the values from the truth table of
$h$ according to
\begin{equation}
\begin{split}
    &\begin{pmatrix} 1 & 0 & 1 & 0 \end{pmatrix} \to h(0,0)\\
    &\begin{pmatrix} 1 & 0 & 0 & 1 \end{pmatrix} \to h(0,1)\\
    &\begin{pmatrix} 0 & 1 & 1 & 0 \end{pmatrix} \to h(1,0)\\
    &\begin{pmatrix} 0 & 1 & 0 & 1 \end{pmatrix} \to h(1,1)\\
\end{split}
\end{equation}
will yield a valid BMP for the function $h(f, g)$. The RTL sweep of
the CLEAN operation is sufficient to compress this BMP.

The transformation from Eq.~(\ref{eq:direct-sum-apply-init}) to
Eq.~(\ref{eq:direct-sum-apply-final1}) proceeds through the following steps.
Initially, the matrix that multiplies the direct sum is given
by
\begin{equation}
    {\bf U}^{(n)} =
    \begin{pmatrix}
        1 & 1
    \end{pmatrix}.
\end{equation}
This matrix is moved to the right via the iterations
\begin{equation}
    {\bf A}^{(k)} =
    \left( \begin{array}{c}
        {\bf U}^{(k)} \;\left[ {\bf F}^{(k-1)}(0) \oplus {\bf G}^{(k-1)}(0) \right] \\
        {\bf U}^{(k)} \;\left[ {\bf F}^{(k-1)}(1) \oplus {\bf G}^{(k-1)}(1) \right]
    \end{array} \right)
    =
    \left( \begin{array}{c}
        {\bf H}^{(k-1)}(0) \\ {\bf H}^{(k-1)}(1)
    \end{array} \right) \cdot
    {\bf U}^{(k-1)}
\end{equation}
that use the Switch-Unique (SU) decomposition (see
Appendix~\ref{SU-decomposition}). The matrices ${\bf U}^{(k)}$ at any
point in the iterations can be decomposed into two row-switching
matrices ${\bf U}^{(k)}_f$ and ${\bf U}^{(k)}_g$ such that
\begin{eqnarray}
    {\bf U}^{(k)} \;\left[ {\bf F}^{(k-1)}(x_{k-1}) \oplus {\bf
        G}^{(k-1)}(x_{k-1}) \right] & = & \left( \begin{array}{c} {\bf
          U}^{(k)}_f \; \; \; {\bf U}^{(k)}_g
    \end{array} \right) \;
    \left[ {\bf F}^{(k-1)}(x_{k-1}) \oplus {\bf G}^{(k-1)}(x_{k-1})
      \right] \\ & = & \left( \begin{array}{cc} {\bf U}^{(k)}_f \;
        {\bf F}^{(k-1)}(x_{k-1}) \;&\; {\bf U}^{(k)}_g \;{\bf
          G}^{(k-1)}(x_{k-1})
    \end{array} \right).
\end{eqnarray}
This allows for an efficient numerical implementation of the procedure
whereby the matrices ${\bf U}^{(k)}$ and ${\bf A}^{(k)}$ are
represented and processed as arrays of tuples (indicating the two
non-zero columns in the $f$ and $g$ blocks), avoiding the explicit
construction of the matrices ${\bf F}^{(k-1)}(x_{k-1}) \oplus {\bf
  G}^{(k-1)}(x_{k-1})$.

Both APPLY methods can be easily extended to handle more than two input
functions. In addition, while it was assumed above that $\bf R$ is the
canonical terminal vector, both methods generalize to the cases where
this is not the case.

%%%%%%%%%%%%%%%%%%%%%%%%%%%%%%%%%%%%%%%%%%%%%%%%%%%%%%%%%%%%%%%%%%%%%%%%%

%%%%%%%%%%%%%%%%%%%%%%%%%%%%%%%%%%%%%%%%%%%%%%%%%%%%%%%%%%%%%%%%%%%%%%%%%
\subsection{RESTRICT}
\label{sec:restrict}

The RESTRICT operation sets a variable $x_i$
of a function to a particular value $b=0$ or $b=1$. Within the BMP
representation, the replacement $x_i \leftarrow b$ is implemented by:
(i) replacing the matrix ${{\bf M}^{(i)}}(x_i)$ in the substring
$\ldots {\bf M}^{(i+1)} (x_{i+1}) \;{\bf M}^{(i)}(x_i) \;{\bf
  M}^{(i-1)}(x_{i-1}) \ldots$ of the full BMP train by ${\bf
  M}^{(i)}(b)$; (ii) absorbing ${\bf M}^{(i)}(b)$ into the right or
left matrix; and (iii) performing the CLEAN operation on the full BMP
matrix train.

%%%%%%%%%%%%%%%%%%%%%%%%%%%%%%%%%%%%%%%%%%%%%%%%%%%%%%%%%%%%%%%%%%%%%%%%%
\subsection{INSERT}
\label{sec:insert}

The INSERT operation adds a muted variable $x_i$ to a Boolean function
with the BMP representation
\begin{equation}
\label{eq:insert}
 f (x_{n - 1}, \ldots, x_{i + 1}, x_{i - 1}, \ldots, x_0) =
 {\bf{F}}^{({n-1})} (x_{n-1}) \cdots {\bf{F}}^{({i-1})}(x_{i-1}) \;
 {\bf{F}}^{({i+1})}(x_{i+1}) \cdots {\bf{F}}^{(0)} (x_{0}) \;{\bf
   R} .
\end{equation}
Inserting a muted (i.e., inactive) variable $x_i$ amounts to adding an
identity matrix of dimensions $d_{i+1}\times d_{i+1}$ at the position
$x_i$ along the train,
\begin{equation}
\label{eq:insert1}
 f (x_{n - 1}, \ldots, x_{i + 1}, x_i, x_{i - 1}, \ldots, x_0) =
 {\bf{F}}^{({n-1})} (x_{n-1}) \cdots {\bf{F}}^{({i-1})}(x_{i-1}) \;
 {\mathbb{1}}^{(i)} \;{\bf{F}}^{({i+1})}(x_{i+1}) \cdots
 {\bf{F}}^{(0)} (x_{0}) \;{\bf R}
 \;,
\end{equation}
where $\mathbb 1$ is the identity matrix.

If the BMP in Eq.~(\ref{eq:insert}) is not in canonical form, the
expanded BMP in Eq.~(\ref{eq:insert1}) must be processed through the
CLEAN operation.

%%%%%%%%%%%%%%%%%%%%%%%%%%%%%%%%%%%%%%%%%%%%%%%%%%%%%%%%%%%%%%%%%%%%%%%%%
\subsection{JOIN}
\label{sec:join}

The JOIN operation combines multiple BMPs representing individual
Boolean functions into a single vector BMP of the type described by
Eq.~(\ref{eq:bmp}). Consider, for example, two functions defined
over the same binary variables, represented by the BMPs shown in
Eq.~(\ref{eq:fg-bmp}). The JOIN operation consists of first
assembling the auxiliary matrices
\begin{equation}
 \label{eq:join}
 {\bf H}^{(i)} (x_i) = \left( \begin{array}{cc} {\bf F}^{(i)} (x_i) &
   0 \\ 0 & {\bf G}^{(i)} (x_i) \\
    \end{array}\right)
\end{equation}
and then building the BMP,
\begin{equation}
\label{eq:join2}
\left(
\begin{array}c
f(x_{n-1},x_{n-2},...,x_1, x_0)\\
g(x_{n-1},x_{n-2},...,x_1, x_0)
\end{array}\right)
= {\bf{H}}^{(n-1)} (x_{n-1}) \;{\bf{H}}^{(n-2)} (x_{n-2}) \cdots
{\bf{H}}^{(1)} (x_1) \;{\bf{H}}^{(0)} (x_0) \;\tilde{\bf R},
\end{equation}
where the terminal vector $\tilde{\bf R} = (0 1 0 1)^\top$. The JOIN
operation then involves: (i) reducing $\tilde{\bf{R}}$ to ${\bf R}$ by
employing a row-switching matrix $\bf{U}$ such that $\tilde{\bf R} =
{\bf U} \;{\bf R}$; (ii) absorbing $\bf{U}$ into the matrix
${\bf{H}}^{(0)} (x_0)$ in Eq.~(\ref{eq:join2}); and (iii) cleaning
the resulting BMP.

When the domains of the functions are not the same, we must work with
the union of domains, for instance, $D = D_f \cup D_g$. For every
missing variable in either $D_f$ or $D_g$, we use INSERT to add
identity matrices of suitable dimensions in the corresponding
BMPs. Moreover, when the BMPs corresponding to different functions do
not have the same variable order, one must align the variable order
using the REORDER operation (see below) before the JOIN operation can
be applied.

%%%%%%%%%%%%%%%%%%%%%%%%%%%%%%%%%%%%%%%%%%%%%%%%%%%%%%%%%%%%%%%%%%%%%%%%%
\subsection{COMPOSE}
\label{sec:compose}

COMPOSE replaces a variable $x_i$ in a Boolean function $f(x_{n-1},
x_{n-2},\dots,x_1,x_0)$ by a Boolean function \newline $g(x_{n-1},
x_{n-2},\dots,x_1, x_0)$. Expressing the Shannon decomposition for the
function \newline $f(x_{n-1}, x_{n-2},\dots, x_i = g({\bf
  x}),\dots,x_1,x_0) = f(x_{n-1}, x_{n-2},\dots, x_i \leftarrow g(x_{n-1},
\dots, x_1,x_0), \dots, x_1,x_0)$ as
\begin{equation}
 f(x_{n-1}, x_{n-2},..., x_i = g({\bf x}),...,x_1,x_0) =
 \left({\overline{g({\bf{x}})}}\; \; g({\bf{x}})\right ) \;\left(
\begin{array}c
f({\bf{x}}; x_i \leftarrow 0)\\
f({\bf{x}}; x_i \leftarrow 1)
\end{array}\right),
\end{equation}
shows that COMPOSE can be easily implemented using the BMP
representation of the functions $f({\bf{x}})$ and $g({\bf{x}})$, and a
combination of the RESTRICT, JOIN, and APPLY operations.

%%%%%%%%%%%%%%%%%%%%%%%%%%%%%%%%%%%%%%%%%%%%%%%%%%%%%%%%%%%%%%%%%%%%%%%%%
\subsection{SWAP}
\label{sec:swap}

The SWAP operation switches the order of two adjacent matrices in the
matrix train. To outline the swap process, consider the product of two
Boolean matrices: $\left[ {\bf M}_2 (x_2) \right]_{n\times p}$ and $
\left[ {\bf M}_1 (x_1) \right]_{p\times m}$,
\begin{equation}
\label{eq:swap}
\left[ {\bf M}_2 (x_2) \right]_{n\times p} \left[ {\bf M}_1(x_1)
  \right]_{p \times m} = \left[ {\bf{M}}_{21}(x_2,x_1)
  \right]_{n\times m}.
\end{equation}
In order to interchange the variables, consider the column-by-column
and row-by-row Shannon decompositions of the matrix ${\bf{M}}_{21}
(x_2, x_1)$ with respect to $x_2$ and $x_1$, respectively, operations
which push $x_1$ to the left and $x_2$ to the right:
\begin{equation}
\label{eq:swap-shannon}
\left[ {\bf{M}}_{21}(x_2,x_1) \right]_{n \times m} = \left[
  {\bf{S}}(x_1) \right]_{n \times 2n} \left[ \bf{\mathcal{M}}
  \right]_{2n\times 2m} \left[ {{\bf{S}}^\top}(x_2) \right]_{2m \times
  m},
\end{equation}
where
\begin{equation}
\left[ \bf{\mathcal{M}} \right]_{2n \times 2m} =
\left[\begin{array}{cc} {\bf{M}}_{21} (0,0) & {\bf{M}}_{21} (1,0)
    \\ {\bf{M}}_{21} (0,1) & {\bf{M}}_{21}
    (1,1) \end{array}\right]_{2n\times 2m} .
\end{equation}
The swapped matrices are constructed from the matrix
$\bf{\mathcal{M}}$ by: (i) eliminating duplicate rows in
$\bf{\mathcal{M}}$ using the RS matrix $\bf{U}$,
\begin{equation}
\label{eq:swap_compression}
\left[\bf{\mathcal{M}}\right]_{2n\times 2m} =
\left[{\bf{U}}\right]_{2n\times r} \left[\tilde{\bf{\mathcal{M}}}
  \right]_{r\times 2m} ,
\end{equation}
with $r\le 2n$; and (ii) splitting $\bf{U}$ into two $n\times r$
blocks and $\tilde{\bf{\mathcal{M}}}$ into two $r\times m$ blocks as
follows:
\begin{equation}
\label{eq:SU_swap_decomposition}
\left[{\bf{U}}\right]_{2n\times r} =
\left[\begin{array}{c}
{\tilde{\bf{M}}_2}(0)\\
{\tilde{\bf{M}}_2}(1)
\end{array}\right]_{2n\times r} , \qquad
\left[\tilde{\bf{\mathcal{M}}}\right]_{r\times 2m} =
\left[{{{\tilde{\bf{M}}}_1} (0)}\;\;{\tilde{\bf{M}}_1}
  (1)\right]_{r\times 2m}.
\end{equation}
From Eqs.~(\ref{eq:shannon_decompositions}),
(\ref{eq:swap-shannon}), (\ref{eq:swap_compression}), and
(\ref{eq:SU_swap_decomposition}) we obtain our final result,
\begin{equation}
{\bf{M}}_{21}(x_2,x_1) = {\bf M}_2 (x_2) \;{\bf M}_1 (x_1) =
{\tilde{\bf{M}}_2} (x_1) \;{\tilde{\bf{M}}_1} (x_2),
\end{equation}
where
\begin{eqnarray}
\tilde{\bf{M}}_2 (x_1) & = & {\bf{S}}(x_1) \;{\bf{U}} =
      {\bf{S}}(x_1) \;\left(\begin{array}{c} {\tilde{\bf{M}}}_2
        (0)\\ {\tilde{\bf{M}}}_2 (1)
\end{array} \right) \\
\tilde{\bf{M}}_1 (x_2) & = & \tilde{\bf{\mathcal{M}}} \cdot
      {\bf{S}}^\top(x_2) = \left( \begin{array}{cc} {\tilde{\bf{M}}_1(0)}
        & {\tilde{\bf{M}}_1}(1) \end{array} \right) \;{\bf{S}}^\top
      (x_1).
\end{eqnarray}
%

%%%%%%%%%%%%%%%%%%%%%%%%%%%%%%%%%%%%%%%%%%%%%%%%%%%%%%%%%%%%%%%%%%%%%%%%%
\subsection{REORDER}
\label{sec:reorder}

The REORDER operation changes the order of the variables by
implementing the appropriate number of SWAP operations.

\subsection{REVERSE ORDER}
\label{sec:invorder}

A particularly simple type of variable reordering operation on BMPs is 
the complete reversal of the original variable
order. This relies on the fact that the transpose of a matrix product
is given by the product of the transposed matrices taken in reverse
order. This case is straightforwardly illustrated for a single Boolean
function $f({\bf x})$, for which we can write:
\begin{align}
{f}(\bf{x})
  &=
  \left[{\bf{M}}^{({n-1})}(x_{n-1})\right]_{1\times m_{n-1}} 
    \cdots \;\left[{\bf{M}}^{(1)}(x_{1})\right]_{m_{2}\times m_{1}} \left[{\bf{M}}^{(0)}
    (x_{0})\right]_{m_{1}\times 2}\;\left[ {\bf R} \right]_{2\times 1}
  \nonumber\\
  &=
  \left[ {\bf R}^\top \right]_{1\times 2}\;
  \left[{\bf{M}}^{(0)}(x_{0})^\top\right]_{2\times m_{1}}\;
  \left[{\bf{M}}^{(1)}(x_{1})^\top\right]_{m_{1}\times m_{2}}
  \cdots\; \left[{\bf{M}}^{({n-1})}(x_{n-1})^\top\right]_{m_{n-1}\times 1}
  \;.
\end{align}
The last line is then brought into row-switching form by first
absorbing the ${\bf R}^\top$ matrix into ${\bf{M}}^{(0)}$, and then
performing the LTR cleaning operation.

A Julia library implementing all operations introduced above has been
developed by us and is publicly available \cite{BMP-library}.
%%%%%%%%%%%%%%%%%%%%%%%%%%%%%%%%%%%%%%%%%%%%%%%%%%%%%%%%%%%%%%%%%%%%%%%%%
\section{Connection to Binary Decision Diagrams}
\label{sec:BDD}

\begin{figure}[h]
    \centering
    \begin{subfigure}{0.5\linewidth}
        \centering
        \includegraphics[scale=1.0,keepaspectratio]{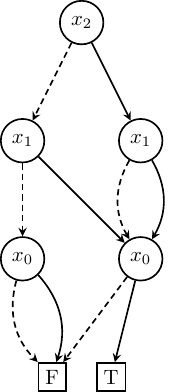}
        \caption{}
        \label{fig:bdd1_a}
    \end{subfigure}%
    \begin{subfigure}{0.5\linewidth}
        \centering
        \includegraphics[scale=1.0,keepaspectratio]{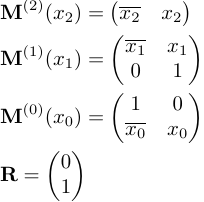}
        \caption{}
        \label{fig:bdd1_b}
    \end{subfigure}
    \caption{(\subref{fig:bdd1_a}) Example of a BDD originated from the BMP in (\subref{fig:bdd1_b}).}
    \label{fig:bdd1}
\end{figure}

BMPs are equivalent to BDDs, namely, for a given a variable order, a
BMP can be transformed into a BDD and vice-versa. (Canonical BMPs are
unique up to local permutation matrix, which amounts to a permutation
of the nodes of a BDD within a level.) The BMP-to-BDD transformation
consists of the following steps. Starting with the BMP
\begin{equation}
    \left[ {\bf M}^{(n-1)}(x_{n-1}) \right]_{p_n \times p_{n-1}}
    \left[ {\bf M}^{(n-2)}(x_{n-2}) \right]_{p_{n-1} \times p_{n-2}}
    \dotsc \left[ {\bf M}^{(1)}(x_{1}) \right]_{p_2 \times p_1} \left[
      {\bf M}^{(0)}(x_{0}) \right]_{p_1 \times p_0} \left[ {\bf R}
      \right]_{p_0 \times 1} ,
\end{equation}
for a site matrix ${\bf M}^{(i)}$, we insert $p_{i+1}$ nodes of
variable $x_i$ to the BDD, each node corresponding to a particular row
of the matrix. We identify these nodes with tuples $(i,a)$ where $a=1,
\dotsc, p_{i+1}$ indicates the row number. We also add $p_0$ terminal
nodes for the elements of the terminal vector ${\bf R}$, identified as
$(-1, a)$, $a = 1, \dotsc, p_0$ ($p_0=2$ for the canonical BMP, but we
keep it general here for notational uniformity). Then, the low and high
children of a node are given by association
\begin{equation}
    \left[ {\bf M}^{(i)}(0) \right]_{ab} = 1 \iff (i-1,b) \text{ is
      the low child of } (i,a)
\end{equation}
and
\begin{equation}
    \left[ {\bf M}^{(i)}(1) \right]_{ab} = 1 \iff (i-1,b) \text{ is
      the high child of } (i,a).
\end{equation}
Following this definition, the children of $x_0$ nodes are the
terminal nodes. In this way, a matrix in the BMP train is equivalent
to the adjacency matrix between two levels of a BDD. As an example,
consider the Boolean function
\begin{equation}
  \label{eq:bdd_ex_f}
    f(x_2, x_1, x_0) = x_0 (x_2 \lor \overline{x}_2 x_1),
\end{equation}
where $\lor$ is the OR operation and integer multiplication represents 
AND.  For the variable order $(x_2, x_1, x_0)$, the BMP of this functions  and the conversion of the BMP to a BDD are shown in
Fig.~\ref{fig:bdd1}.

%\end{document}

%%%%%%%%%%%%%%%%%%%%%%%%%%%%%%%%%%%%%%%%%%%%%%%%%%%%%%%
%%%%%%%%%%%%%%%%%%%%%%%%%%%%%%%%%%%%%%%%%%%%%%%%%%%%%%%

\begin{figure}
  \centering
  \begin{subfigure}{0.32\textwidth}
    \includegraphics[scale=0.75,keepaspectratio]{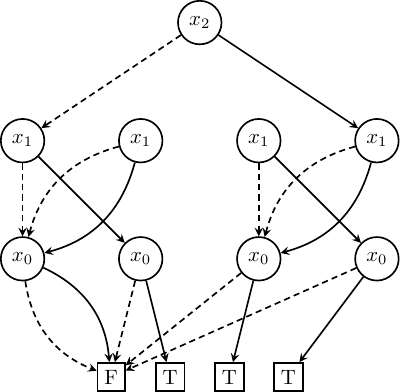}
    \caption{}
    \label{fig:bdd_clean:a}
  \end{subfigure}
  \hfill
  \begin{subfigure}{0.32\textwidth}
    \includegraphics[scale=0.75,keepaspectratio]{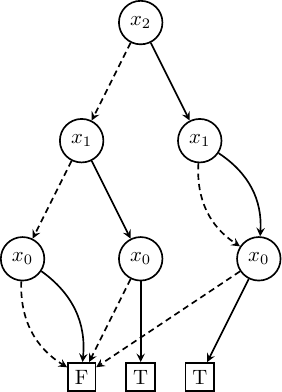}
    \caption{}
    \label{fig:bdd_clean:b}
  \end{subfigure}
  \hfill
  \begin{subfigure}{0.32\textwidth}
    \includegraphics[scale=0.75,keepaspectratio]{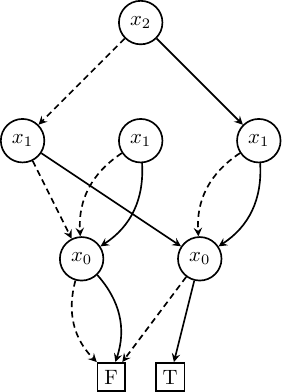}
    \caption{}
    \label{fig:bdd_clean:c}
  \end{subfigure}
  \caption{(\subref{fig:bdd_clean:a}) The BDD representation of the BMP in
    Eq.~(\ref{eq:BMP-example}). (\subref{fig:bdd_clean:b}) After LTR cleaning. (\subref{fig:bdd_clean:c}) After RTL
    cleaning.}
    \label{fig:bdd_clean}
\end{figure}

We note that going left to right on the BMP corresponds to moving top
to bottom on the equivalent BDD. This correspondence allows one to
visualize what the LTR and RTL cleaning sweeps accomplish. The former
removes the disconnected subgraphs, while the latter merges isomorphic
ones. To justify these statements consider the function in
Eq.~(\ref{eq:bdd_ex_f}), which can be rewritten in the form $f({\bf
  x}) = g({\bf x}) \lor h({\bf x})$, where $g({\bf x}) = x_2 x_0$ and
$h({\bf x}) = \overline{x}_2 x_1 x_0$ (this function $h$ was used as
the example in Sec.~\ref{sec:example}). The BMPs representing these functions are given by
\begin{equation}
    g(x_2,x_1,x_0) =
    \begin{pmatrix}
        \overline{x}_2 & x_2
    \end{pmatrix}
    \begin{pmatrix}
        1 & 0 \\ 0 & 1
    \end{pmatrix}
    \begin{pmatrix}
        1 & 0 \\ \overline{x}_0 & x_0
    \end{pmatrix}
    \begin{pmatrix}
        0 \\ 1
    \end{pmatrix}
\end{equation}
and
\begin{equation}
    h(x_2,x_1,x_0) =
    \begin{pmatrix}
        \overline{x}_2 & x_2
    \end{pmatrix}
    \begin{pmatrix}
        \overline{x}_1 & x_1 \\ 1 & 0
    \end{pmatrix}
    \begin{pmatrix}
        1 & 0 \\ \overline{x}_0 & x_0
    \end{pmatrix}
    \begin{pmatrix}
        0 \\ 1
    \end{pmatrix}.
\end{equation}
If $f$ is constructed from the BMPs of $g$ and $h$ via APPLY using the
direct product method, the following BMP is generated before
cleaning:
\begin{equation}
    f({\bf x}) =
    \begin{pmatrix}
        \overline{x}_2 & 0 & 0 & x_2
    \end{pmatrix}
    \begin{pmatrix}
        \overline{x}_1 & x_1 & 0 & 0\\
        1 & 0 & 0 & 0\\
        0 & 0 & \overline{x}_1 & x_1\\
        0 & 0 & 1 & 0
    \end{pmatrix}
    \begin{pmatrix}
        1 & 0 & 0 & 0\\
        \overline{x}_0 & x_0 & 0 & 0\\
        \overline{x}_0 & 0 & x_0 & 0\\
        \overline{x}_0 & 0 & 0 & x_0
    \end{pmatrix}
    \begin{pmatrix}
        0 \\ 1 \\ 1 \\ 1
    \end{pmatrix}.
    \label{eq:BMP-example}
\end{equation}
The BDD corresponding to this intermediate (non-canonical) BMP is shown
in Fig.~\ref{fig:bdd_clean:a}. Bringing this BMP to canonical form
requires applying the CLEAN operation in both directions. Single
sweeps of LTR and RTL CLEAN produce the BDDs shown in
Figs.~\ref{fig:bdd_clean:b} and \ref{fig:bdd_clean:c},
respectively. Clearly, further compression of 
the BDD in Fig.~\ref{fig:bdd_clean:c} is possible. This is accomplished through an additional LTR CLEAN sweep, which, in turn, leads to the BDD shown in Fig.~\ref{fig:bdd1_a}.

%%%%%%%%%%%%%%%%%%%%%%%%%%%%%%%%%%%%%%%%%%%%%%%%%%%%%%%
%%%%%%%%%%%%%%%%%%%%%%%%%%%%%%%%%%%%%%%%%%%%%%%%%%%%%%%

The translation from BDDs to BMPs is not as straightforward: BDDs allow
for nodes at the level $l$ to connect to nodes at levels $j > l+1$
(i.e., at non consecutive levels). This is not the case for BMPs and
thus, the translation to a BMP requires that one first pads the BDD
with pass-through nodes to ensure that all nodes in a level are only
connected to nodes in the next level. With this padding in place, the
matrices in the BMP correspond to the adjacency matrices between the
levels of the BDD.

The padded BDDs described above, which involve only connections
between consecutive layers, are referred to as
\textit{complete}. Complete BDDs that are further simplified by
merging equivalent nodes are referred to as \textit{quasi-reduced}. Fully
compressed BMPs correspond to quasi-reduced BDDs, a type of BDDs used in the parallelization of BDD
operations \cite{Ochi1991, Ochi1993}. While the bond dimension across
the BMP and the number of nodes in the BDD scale similarly with the number of variables, the
the bypassing of nodes in non-complete BDDs 
often makes those BDDs smaller data structures than BMPs. For that same
reason, BDDs can be faster to evaluate than BMPs (by evaluation we
mean the computation of the output $f({\bf x})$ given the input ${\bf
  x}$). We note that both BDDs and BMPs can be evaluated by traversal:
for BDDs, traversal of the tree, always from root to terminal node;
for BMPs, traversal of the matrix train. However, BMPs offer additional forms of evaluation, including
left-to-right, right-to-left, and other forms of matrix train
contraction, including parallelized decimation.

%%%%%%%%%%%%%%%%%%%%%%%%%%%%%%%%%%%%%%%%%%%%%%%%%%%%%%%%%%%%%%%%%%%%%%%%%
\section{Computational complexity and optimization}
\label{sec:complexity}

The computational complexity of BMPs is diagnosed through the scaling
of the accumulated bond dimension or, interchangeably, of the maximum
bond dimension along the matrix train with the number of variables
$n$. For products of row-switching matrices the accumulated bond
dimension measures the combined number of matrix elements -- i.e., the
total number of rows -- across all matrices of the BMP, a measure
which, when translated to BDDs, corresponds to the number of
nodes. For brevity, hereafter the accumulated bond dimension across
the BMP matrix train and the number of nodes of a BDD will be referred
to as the BMP- and BDD-volume, respectively. For an arbitrary order of
variables, a generic Boolean function has exponential complexity,
i.e., its volume scales exponentially with the number of variables,
$n$. Generally, efficient generation/manipulation of BMPs and BDDs
refers to cases for which one can find a variable ordering for which
the volume of the BMP or BDD scales polynomially with $n$. In
practice, both BMPs and BDDs remain useful as long as their volumes
are kept computationally manageable, even when the scaling with $n$ is
exponential. Below, we discuss the ordering problem in the context of
BMPs and provide both ``exact" and heuristic methods for optimizing
the BMP volume via ordering protocols. While both methods are
presented in the context of a fully developed BMP, in practice, one or
a combination of variable reordering algorithms can be implemented in
the course of building the BMP as a way of controlling the volume once
the volume increases to a threshold value. This section also discusses
the difference in complexity between the direct-product and direct-sum
implementations of the BMP APPLY operation.

%%%%%%%%%%%%%%%%%%%%%%%%%%%%%%%%%%%%%%%%%%%%%%%%%%%%%%%%%%%%%%%%%%%%%%%%%
\subsection{The variable order problem}
\label{sec:variable-order}

Even though for a generic Boolean function with an arbitrary variable
order the BMP volume scales exponentially with $n$, many commonly
encountered Boolean functions display polynomial-volume BMPs provided
that a suitable variable order is found. As an example, consider the
Boolean function
\begin{equation}
    f({\bf x}) = x_0\, x_1 \lor x_2\, x_3 \lor \dotsc \lor x_{2k-2}\, x_{2k-1}
\end{equation}
of $n = 2k$ input variables, where $\lor$ is the logical OR operation.
Initially, let us assume that the corresponding BMP is constructed
according to a variable order in which all the even variables precede
all the odd variables, i.e., $x_0, x_2, \dotsc, x_{2k-2}, x_1, x_3,
\dotsc, x_{2k-1}$. The bond dimension of the $l$-th matrix, $p_{l}$,
is equal to the number of distinct functions obtained upon fixing the
value of the first $l$ variables in the variable order. For $l\leq k$,
we set $x_{j} = v_{j}$ for $j=0,\cdots,2l-2$. Then, the subfunctions
are of the form
\begin{equation}
  v_0\, x_1 \lor v_2\, x_3 \lor \cdots \lor v_{2l-2}\, x_{2l-1}
  \lor x_{2l}\, x_{2l+1} \lor \cdots \lor x_{2k-2}\, x_{2k-1},
\end{equation}
which are clearly distinct for any assignment of values $v_0, v_2,
\dotsc, v_{2l-2}$. Therefore, $p_{l} = 2^l$ in this part of the BMP
train. ($p_{0} = 1$, as expected.) For $l = k + m > k$, in addition to
fixing the value of all even variables, we also fix the value of the
first $m$ odd variables, obtaining subfunctions of the form
\begin{equation}
    v_0\, v_1 \lor v_2\, v_3 \lor \dotsc \lor v_{2m-2}\, v_{2m-1} \lor v_{2m}\,
    x_{2m+1} \lor \cdots \lor v_{2k-2}\, x_{2k-1}.
\end{equation}
Notice that if any of the constant terms evaluate to $1$, then the
entire function is identically $1$. Otherwise, each distinct
assignment of values $v_{2m}, \cdots, v_{2k-2}$ leads to a distinct
function, resulting in a bond dimension $p_{k+m} = 1+2^{k-m}$.
(Notice that for $m = k$ or, equivalently, $l = 2k = n$, we have
$p_{2k} = 2$, which is the dimension of the canonical terminal vector
$R$. We do not include the $R$ vector in the volume of the BMP.)  The
BMP volume is then
\begin{equation}
    \sum_{l=0}^k 2^l + \sum_{m=1}^{k-1} \left( 1 + 2^{k-m} \right) =
    2^{k+1} - 1 + (k-1) + 2 \left( 2^{k-1} - 1 \right) = 3\cdot2^{n/2}
    + n/2 - 4.
\end{equation}

Now assume instead that the variable order follows the variable
indices, i.e., $x_0, x_1, \cdots, x_{2k-2}, x_{2k-1}$. For odd $l >
1$, the subfunctions are of the form
\begin{equation}
    v_0\, v_1 \lor v_2\, v_3 \lor \cdots \lor v_{l-1}\, x_{l} \lor x_{l+1}\,
    x_{l+2} \lor \cdots \lor x_{2k-2}\, x_{2k-1}.
\end{equation}
Such a subfunction is identically $1$ if the constant terms evaluate
to $1$. Otherwise, it has two distinct forms depending on the value
$v_{l-1}$. Therefore, all odd bonds have dimension $3$, except for
$l=1$, for which the constant term is absent and the bond dimension is
then $2$. For all even $l$, the subfunctions have the form
\begin{equation}
    v_0\, v_1 \lor v_2\, v_3 \lor \cdots \lor v_{l}\, v_{l+1} \lor x_{l+2}\,
    x_{l+3} \lor \cdots \lor x_{2k-2}\, x_{2k-1},
\end{equation}
which is either a constant equal to $1$ or the sum of the non-constant
terms, depending on whether or not the constant terms add up to $1$. The
total volume (excluding the terminal vector) is then given by
\begin{equation}
    1 + 2 + 2 + \sum_{m=2}^{k-1} (3 + 2) + 3 = 5k - 2 = 5n/2 - 2.
\end{equation}
In this example, we see that while a bad choice of the variable order
results in the BMP volume growing exponentially in the number of variables,
there exists an order that yields a linear-size BMP. 

It should be noted that in most cases of practical interest, we cannot
work with the explicit formula of the Boolean function to compute the
volume of its BMP for a particular variable order, or even find the
optimal one, although methods exist to do so for specific classes of
functions. As outlined in the introduction to this section, here we
adapt two families of variable ordering optimization algorithms
proposed for and tested on BDDs: one ``exact" but of exponential
complexity and thus of more limited use, and the other heuristic but
broadly applied to problems of practical interest.

%%%%%%%%%%%%%%%%%%%%%%%%%%%%%%%%%%%%%%%%%%%%%%%%%%%%%%%%%%%%%%%%%%%%%%%%%
\subsubsection{Exact minimization}
\label{sec:exact-min}

The exact minimization algorithm formulates
the problem of finding the optimal variable ordering as a search in a
state space \cite{Drechsler2004}. In this formulation, a state $q$ is
a subset of the input variables $X_n = \{x_1, x_2, \dotsc, x_n\}$,
representing all BMPs whose first $|q|$ variables are those in $q$,
\textit{in any order}. This state can transition into states $q^\prime
= q \cup \{x_i\}$ such that $x_i \notin q$, i.e., to those states with
an extra variable. Therefore, this state space corresponds to a
directed graph whose links are associated with the individual input
variables of the BMP. In addition, a \textit{path} from the state
$\emptyset$ to state $q \subset X_n$ describes a particular ordering
of the first $|q|$ variables of the BMP. For example, with five
variables, one such path would be
\begin{equation}
  \label{eq:example_path1}
    \emptyset \xrightarrow{x_2} \{x_2\} \xrightarrow{x_4} \{x_2, x_4\}
    \xrightarrow{x_1} \{x_1, x_2, x_4\} \xrightarrow{x_5} \{x_1, x_2,
    x_4, x_5\} \xrightarrow{x_3} \{x_1, x_2, x_3, x_4, x_5\},
\end{equation}
which corresponds to the variable order $(x_2, x_4, x_1, x_5,
x_3)$. Another possible path is
\begin{equation}
  \label{eq:example_path2}
    \emptyset \xrightarrow{x_4} \{x_4\} \xrightarrow{x_1} \{x_1, x_4\}
    \xrightarrow{x_2} \{x_1, x_2, x_4\} \xrightarrow{x_3} \{x_1, x_2,
    x_3, x_4\} \xrightarrow{x_5} \{x_1, x_2, x_3, x_4, x_5\}
\end{equation}
corresponding to the ordering $(x_4, x_1, x_2, x_3, x_5)$. Notice that
the third state on the paths is the same for both paths, but is
reached via different routes.

The goal now is to construct a path from $\emptyset$ to $X_n$ that
gives the optimal variable ordering. To express this as a
shortest-path problem, we define a cost function $c(q,q^\prime)$ for
all transitions such that their sum along such a path is the total
volume of the BMP with the prescribed variable ordering. Since each
transition is associated with a variable, we can simply take $c(q, q
\cup \{x_i\})$ to be the number of rows of the matrix for $x_i$,
subject to the condition that the first $|q|$ variables are given by the set
$q$, and the $(|q|+1)$-th variable is $x_i$. Note that, as can be seen
from the Shannon decomposition, this choice does not depend on the
ordering of the first $|q|$ variables, or on the variables after
$x_i$, and thus the cost function can be defined consistently, independent of
the specific path between $q$ and $q \cup \{x_i\}$. The total
cost of a path from $\emptyset$ to $X_n$ is then indeed just the sum of
the number of rows of all matrices, i.e., the
BMP volume.

While the shortest path can be found using a variety of means,
following \cite{Drechsler2004}, we employ
A*~\cite{hart1968formal}. With A*, two maps are constructed by the
algorithm: $g(q)$, the cost of the currently known shortest path to
$q$, and $h(q)$, a (lower-bound) estimate of the cost from $q$ to the
target state (in our case, $X_n$). Furthermore, a set of ``open'' states
is maintained. In each iteration, the state with the smallest value of
$g(q) + h(q)$ is chosen from this set for processing. (This way, the
algorithm prioritizes the most promising states.) The processing
involves updating the values $g(q^\prime)$ for the successor states
$q^\prime$ of $q$ according to
\begin{equation}
    g(q^\prime) \gets \min \big( g(q^\prime), g(q) + c(q,q^\prime)
    \big),
\end{equation}
and inserting them to the open states set if the cost is changed. If
$q$ is the target state, the search is ended, with $g(q)$ being the
cost associated with the shortest path. If, in addition to $g$, the
predecessor of each state $q$ is stored during the iterations, the
shortest path itself can be reconstructed as well.

A* uses a heuristic in the form of $h(q)$, but it works exactly so
long as this function is a lower bound on the actual cost from $q$ to
the target state. For the purpose of finding the optimal variable
ordering, $h(q)$ needs to be a lower bound on the sum of the
dimensions of the last $n-|q|$ matrices of a BMP, where the first
$|q|$ matrices belong to the variables in $q$. Consider such a BMP
where the $|q|+1$-th variable is $x_i$. As remarked earlier, the
number of rows of the matrix for $x_i$ does not depend on the ordering
of the first $|q|$ variables, but on top of that, it does not depend
on $x_i$ either. Let this number be given by a function
$\chi(q)$. (With this definition, $c(q, q \cup \{x_i\}) = \chi(q)$,
i.e., the distance from a state to any of its successors is the same.)
Then, a simple choice for $h(q)$ is
\begin{equation}
    h(q) = \chi(q) + (n-|q|-1).
\end{equation}
$\chi(q)$ can be computed by constructing any variable order
represented by $q$ via local SWAPs and looking at the dimension of the
$|q|+1$-th matrix.

In \cite{Drechsler2004}, the approach based on A* is further improved
by combining it with branch-and-bound techniques. In the context of
BMPs, when the search algorithm is processing a state $q$, it has to
reconstruct the BMP that corresponds to this state in order to compute
and estimate the quantities mentioned earlier. This is usually done by
locally swapping the variables in the BMP until a suitable ordering is
reached. This means that as A* is running, various full variable
orderings need to be constructed even though these are not relevant to
the A* algorithm. By keeping track of the size of these full variable
orderings, the shortest path can be bounded from above. This extra
information can help prune the search space. If the estimated length
of a path is greater than this upper bound it does not need to be
considered at all. This might be useful in keeping the open states
queue small. Furthermore, if there is a point when the lower bound
computed by A* and the upper bound found while performing swaps
coincide, the algorithm can be terminated early. The only possibility
in such a case is that the variable order that gives the upper bound
is the optimal one.

%%%%%%%%%%%%%%%%%%%%%%%%%%%%%%%%%%%%%%%%%%%%%%%%%%%%%%%%%%%%%%%%%%%%%%%%%
\subsubsection{Heuristic methods}
\label{sec:heuristic}

In many cases, simpler heuristic methods for improving the variable
order suffice. These methods are implemented on given BMPs via the
SWAP operation, which changes the order of two adjacent variables in
the matrix train.

A simple but highly efficient method is the sifting algorithm proposed
by Rudell in \cite{Rudell1993} for BDDs, which can be readily adapted
for BMPs. With sifting, a single variable is moved around along the
train using SWAP while the relative ordering of all the other
variables is left unchanged. After the optimal position for the one
variable is determined, it is moved to that site using SWAP
again. With five variables, this may look like
\begin{align*}
    & x_1, \boldsymbol{x_2}, x_3, x_4, x_5\\
    & \boldsymbol{x_2}, x_1, x_3, x_4, x_5\\
    & x_1, \boldsymbol{x_2}, x_3, x_4, x_5\\
    & x_1, x_3, \boldsymbol{x_2}, x_4, x_5\\
    & x_1, x_3, x_4, \boldsymbol{x_2}, x_5\\
    & x_1, x_3, x_4, x_5, \boldsymbol{x_2}
\end{align*}
followed by
\begin{align*}
    & x_1, x_3, x_4, x_5, \boldsymbol{x_2}\\
    & x_1, x_3, x_4, \boldsymbol{x_2}, x_5\\
    & x_1, x_3, \boldsymbol{x_2}, x_4, x_5
\end{align*}
if $x_2$ is the variable being ``sifted'' and the optimal position for
it is the third site. This process is to be repeated for all the
variables, possibly multiple times.

Because of the bounds on the approximation of good variable order for
BMPs, the method is not guaranteed to find a small enough BMP, but it
works well for large classes of functions in practice, especially when
compared to other methods.

%%%%%%%%%%%%%%%%%%%%%%%%%%%%%%%%%%%%%%%%%%%%%%%%%%%%%%%%%%%%%%%%%%%%%%%%%
\subsubsection{Comparison of reordering schemes}
\label{sec:comparison}

\begin{table}
    \small
    \centering
    \begin{tabular}{c|c|c|c|c|c|c}
        $n$ & Initial size & A* runtime & A* + B\&B runtime & Minimum BMP size & Sifting runtime & Sifting output size \\
        \hline
        $8$ & $2766$ & $1.751811$ s & $1.851496$ s & $151$ & $0.002581$ s & $222$\\
        $10$ & $11204$ & $49.684622$ s & $49.071847$ s & $211$ & $0.020537$ s & $329$\\
        $12$ & $44986$ & $235.863001$ s & $230.140865$ s & $279$ & $0.045474$ s & $456$\\
        $14$ & $180144$ & $6173.582738$ s & $5593.707498$ s & $355$ & $0.291017$ s & $603$
    \end{tabular}
    \caption{Comparison of results for different variable ordering
      algorithms for a BMP representing an $n$-bit full adder.}
    \label{tab:var-order}
\end{table}

Table \ref{tab:var-order} shows the runtimes and resulting BMP volumes
for three different methods of volume minimization via variable
reordering for a BMP implementing a full adder circuit
\cite{Taub1982}, with $n$ being the number of bits of each
operand. (Therefore, the circuit has $2n$ input bits and
$n+1$ output bits.) All BMP manipulations are
performed using the Julia library developed by the authors
\cite{BMP-library}. The minimal (i.e., optimal) BMP volume is obtained
with both A* and with a combination of A* and branch-and-bound techniques. The latter becomes more efficient as $n$ increases. However, when compared to the A* based exact optimization schemes, sifting is
several orders of magnitudes faster.

\subsection{Complexity of APPLY methods}
\label{sec:complexity-apply}

APPLY is the most frequently used operation in the synthesis of BMPs
from logic circuits or Boolean functions. Here we explain the
underlying reason for the differences in computational complexity of
the two APPLY methods. For this purpose, let us explicitly keep track of
the dimensions of the matrices in Eq.~(\ref{eq:fg-bmp}):
\begin{equation}
    f({\bf x}) = \left[ {\bf F}^{(n-1)}(x_{n-1}) \right]_{p_{n}\times
        p_{n-1}} \left[ {\bf F}^{(n-2)}(x_{n-2}) \right]_{p_{n-1}\times
        p_{n-2}} \dotsc \left[ {\bf F}^{(1)}(x_{1}) \right]_{p_{2}\times
        p_{1}} \left[ {\bf F}^{(0)}(x_{0}) \right]_{p_{1}\times p_{0}}
    \left[ {\bf R} \right]_{p_0 \times 1}
\end{equation}
and
\begin{equation}
    g({\bf x}) = \left[ {\bf G}^{(n-1)}(x_{n-1}) \right]_{q_{n}\times
        q_{n-1}} \left[ {\bf G}^{(n-2)}(x_{n-2}) \right]_{q_{n-1}\times
        q_{n-2}} \dotsc \left[ {\bf G}^{(1)}(x_{1}) \right]_{q_{2}\times
        q_{1}} \left[ {\bf G}^{(0)}(x_{0}) \right]_{q_{1}\times q_{0}}
    \left[ {\bf R} \right]_{q_0 \times 1}.
\end{equation}
The BMP representing the Boolean function $h(f({\bf x}),g({\bf x}))$
before cleaning has the form
\begin{equation}
    \left[ {\bf H}^{(n-1)}(x_{n-1}) \right]_{r_{n}\times r_{n-1}}
    \left[ {\bf H}^{(n-2)}(x_{n-2}) \right]_{r_{n-1}\times r_{n-2}}
    \dotsc \left[ {\bf H}^{(1)}(x_{1}) \right]_{r_{2}\times r_{1}}
    \left[ {\bf H}^{(0)}(x_{0}) \right]_{r_{1}\times r_{0}} \left[ {\bf
        R} \right]_{r_0 \times 1},
\end{equation}
where $r_i = p_i q_i$ for the direct product method and $r_i \leq p_i
q_i$ for the direct sum method. The complexity of the operation is
determined by the size of these matrices generated in the intermediate
stage and differs for each method.

Let ${\bf v}_i$ and ${\bf x}^\prime_i$ denote binary vectors of length
$n-i$ and $i$, respectively. Consider subfunctions $f({\bf v}_i, {\bf
    x}^\prime_i)$ of $f$, defined on the subset of the input variables
$x_{i-1}, \dotsc, x_0$ obtained upon fixing the values of $x_{n-1},
\dotsc, x_i$ to ${\bf v}_i$. The BMP of $f$ tells us that these
subfunctions come from a collection of $p_i$ functions on $x_i,
\dotsc, x_0$, which we can label $f_{i,a}$ with $a = 1, \dotsc,
p_i$. (If the BMP has not been compressed via CLEAN, some of these
functions may not be in the Shannon
expansion of $f$, or if they are, they may be redundant.) In other words, for some ${\bf v}_i \in
\{0,1\}^{n-i}$
\begin{equation}
    f\left( {\bf v}_i, \cdot\right) \in \{ f_{i,1}, f_{i,2}, \dotsc,
    f_{i, p_i} \}.
\end{equation}
Similarly, we can write
\begin{equation}
    g\left( {\bf v}_i, \cdot\right) \in \{ g_{i,1}, g_{i,2}, \dotsc,
    g_{i, q_i} \}.
\end{equation}
The purpose of APPLY is to obtain the same set for $h$. The direct
product method does this by generating all $p_i q_i$ combinations of
the form
\begin{equation}\label{eq:direct_prod_comb}
    h\left( f_{i,a}({\bf x}_i), g_{i,b} ({\bf x}_i) \right).
\end{equation}
However, not all of these are subfunctions of $h$, which must be of
the form
\begin{equation}
    h\left( f({\bf v}_i, {\bf x}^\prime_i), g({\bf v}_i, {\bf
        x}^\prime_i) \right).
\end{equation}
Some of the subfunctions in Eq.~(\ref{eq:direct_prod_comb}) are
unnecessary, and only combinations $h(f_{i,a}, g_{i,b})$
corresponding to the same value of ${\bf v}_i$ are relevant. This is
precisely why the additional LTR CLEAN is needed in optimally compressing the BMP resulting from the direct product
method. The direct sum fixes this issue by eliminating the unwanted subfunctions and  
only keeps the relevant combinations, encoded in the matrix ${\bf
  U}^{(i)}$. At any point during the iterative process, ${\bf
  U}^{(i)}$ is a concatenation of two row-switching matrices,
\begin{equation}
    {\bf U}^{(i)} =
    \begin{pmatrix}
        {\bf U}^{(i)}_f & {\bf U}^{(i)}_g
    \end{pmatrix}.
\end{equation}
This matrix can be viewed as an array of pairs, $u^{(i)}_k = (a_k,
b_k)$, $a_k$ and $b_k$ being the non-zero columns on the $k$-th row in
the $f$ and $g$ blocks, respectively. This array enumerates the
combinations $h(f_{i,a_k}, g_{i,b_k})$ that show up in the Shannon
expansion of $h$ at this order. The next order is obtained by
constructing
\begin{equation}
    {\bf A}^{(i)} = \begin{pmatrix}
        {\bf U}^{(i)}_f {\bf F}^{(i)}(0) & {\bf U}^{(i)}_g {\bf G}^{(i)}(0) \\
        {\bf U}^{(i)}_f {\bf F}^{(i)}(1) & {\bf U}^{(i)}_g {\bf G}^{(i)}(1)
    \end{pmatrix}.
\end{equation}
The upper half of this matrix enumerates the combinations
\begin{equation}
    h\left( f_{i,a_k}(x_{i-1}=0, \cdot), g_{i,b_k}(x_{i-1}=0, \cdot)
    \right) = h\left( f_{i-1,a^\prime_k}(\cdot), g_{i-1, b^\prime_k}
    (\cdot) \right),
\end{equation}
consistent with
\begin{equation}
    \left[ {\bf F}^{(i-1)}(x_{i-1} = 0) \right]_{a_k,j} = \delta_{j,
        a^\prime_k},\quad \left[ {\bf G}^{(i-1)}(x_{i-1} = 0)
    \right]_{b_k,j} = \delta_{j, b^\prime_k}.
\end{equation}
The same relations hold true for the $x_{i-1}=1$ substitution as well,
and the corresponding matrix products form the lower half of ${\bf
  A}^{(i)}$. Note that the number of rows of ${\bf H}^{(i)}$, $r_n$,
is the same as for ${\bf U}^{(i)}$.  Since ${\bf A}^{(i)}$ has twice as
many rows as ${\bf U}^{(i)}$ and its unique rows make up ${\bf
  U}^{(i-1)}$, we have the bound
\begin{equation}
    r_{i-1} \leq 2 r_i.
\end{equation}
Furthermore, because there are only $p_{i-1}$ and $q_{i-1}$ columns
in the $f$ and $g$ blocks of ${\bf A}^{(i)}$, respectively, the number of unique
rows is also limited by
\begin{equation}
    r_{i-1} \leq p_{i-1} q_{i-1}.
\end{equation}
The direct product method always saturates the second bound but, in many cases,
violates the first one. This explains the advantage of the
direct sum method for generic Boolean functions, especially when the bond dimensions of input BMPs are large.

%%%%%%%%%%%%%%%%%%%%%%%%%%%%%%%%%%%%%%%%%%%%%%%%%%%%%%%%%%%%%%%%%%%%%%%%%
\section{Summary and Discussion}\label{sec:conclusions}
We have introduced a novel representation of Boolean functions that
employs a train of binary matrices that we refer to as Boolean matrix
products (BMPs), in analogy with the matrix product states used to
represent quantum mechanical many-body states. We showed that, when
maximally compressed, BMPs are normal forms; and introduced operations
on BMPs that allows one to synthesize, manipulate, and combine
BMPs. BMPs are closely related to ordered binary decision diagrams
(BDDs), a connection we made explicit by providing the translation
between the two Boolean function representations. Finally, we
discussed the complexity of BMPs, reflected in the maximum bond
dimension (i.e., the maximum number of rows of any of the matrices
participating in the matrix train), which we associate to the their
total volume. Similarly with the volume of BDDs (i.e., the total
number of nodes of the BDD) the bond dimension of a BMP is highly
sensitive to variable order, an issue considered in this paper in the
context of two approaches to variable reordering -- one exact and
another heuristic.

While we have not identified a ``killer application" that demonstrates
a strong benefit of the BMPs over the BDDs, our motivation for
introducing BMPs was practical: the Colorado University Decision
Diagram (CUDD) package developed by Fabio
Somenzi~\cite{somenzi1998cudd} in the late 1990s is no longer fully
supported and other emerging platforms for manipulating
BDDs~\footnote{See, for example,
    https://github.com/johnyf/tool\_lists/blob/main/bdd.} -- many of which
rely on elements of CUDD -- are difficult to employ or modify for
specific use cases. The principle advantage of BMPs is that their
evaluation and manipulation only requires linear algebra tools, a
simplifying feature that should prompt the development of open-source
libraries that are more flexible, easier to use, and easier to modify
than those currently available for BDDs. Our first implementation of a
GitHub BMP library can be found at~\cite{BMP-library}. The direct
analogy with tensor-train and tensor network schemes used in numerical
analysis and many-body physics should help expand the interest in and
use of these techniques to the mathematics and physics communities
where the BMP representation should find applications to random
classical circuits and to the simulation and verification of quantum
circuits (see, for example, references describing the application of decision diagram techniques to quantum circuits, ~\cite{miller2006qmdd,niemann2015qmdds,hong2020tensor,zhang2024quantum,Hillmich2024Efficient}).

\pagebreak

%%%%%%%%%%%%%%%%%%%%%%%%%%%%%%%%%%%%%%%%%%%%%%%%%%%%%%%%%%%%%%%%%%%%%%%%%
%%%%%%%%%%%%%%%%%%%%%%%%%%%%%%%%%%%%%%%%%%%%%%%%%%%%%%%%%%%%%%%%%%%%%%%%%

\section{Appendix}
\label{sec:Appendix}

This appendix summarizes some of the details needed in the
construction and evaluation of BMPs.

%%%%%%%%%%%%%%%%%%%%%%%%%%%%%%%%%%%%%%%%%%%%%%%%%%%%%%%%%%%%%%%%%%%%%%%%%
\subsection*{Row-switching matrices}

The matrices that comprise a BMP are row switching, that is, all
entries in a row are zero except one that equals $1$. We can associate
with each such matrix ${\bf M}$ of size $p\times q$, an array $m$ of length
$p$ containing values in the range $1, \dotsc, q$, such that
\begin{equation}
    \left[ {\bf M} \right]_{ij} = \delta_{j, m[i]}.
\end{equation}
Furthermore, we can perform various matrix operations using $m$
without needing to construct ${\bf M}$ explicitly. Thus, BMPs can be
implemented entirely in terms of arrays $m$, saving both time and
space. In this appendix, we describe the algorithms we use for this
purpose.

As already noted in the body of the paper, matrices in the BMP can be interpreted as adjacency
matrices between two levels. The array representation described here
is effectively an \textit{adjacency list} representation, which is
efficient because the number of children of each node is fixed at two.

%%%%%%%%%%%%%%%%%%%%%%%%%%%%%%%%%%%%%%%%%%%%%%%%%%%%%%%%%%%%%%%%%%%%%%%%%
\subsection*{Simple product}

Assume we have row-switching matrices $\left[{\bf A}\right]_{p\times q}$ and
$\left[{\bf B}\right]_{q\times r}$. Let $a$ and $b$ be the arrays associated
with these matrices. If ${\bf C} = {\bf A} \; {\bf B}$, we have
\begin{equation}
\begin{split}
    \left[{\bf C}\right]_{ij} &= \sum_{k=1}^{q} \left[{\bf A}\right]_{ik} \left[{\bf B}\right]_{kj}\\
    &= \sum_{k=1}^{q} \delta_{k,a[i]} \delta_{j, b[k]}\\
    &= \delta_{j, b[a[i]]}.
\end{split}
\end{equation}
We see that the matrix ${\bf C}$ is then associated with an array $c$ such
that $c[i] = b[a[i]]$.

%%%%%%%%%%%%%%%%%%%%%%%%%%%%%%%%%%%%%%%%%%%%%%%%%%%%%%%%%%%%%%%%%%%%%%%%%
\subsection*{Direct product}

Let matrices $\left[{\bf A}\right]_{p\times q}$ and $\left[{\bf B}\right]_{r\times
  s}$ be given by arrays $a$ and $b$ once again. Their direct product
$\left[{\bf C}\right]_{pr\times qs} = {\bf A} \otimes {\bf B}$ has entries
\begin{equation}
\begin{split}
    \left[{\bf C}\right]_{(i-1)p+k, (j-1)q+l} &= \left[{\bf A}\right]_{ij} \left[{\bf B}\right]_{kl}\\
    &= \delta_{j, a[i]} \delta_{l, b[k]}.
\end{split}
\end{equation}
Defining $I = (i-1)p+k$ and $J = (j-1)q+l$, we can write
\begin{equation}
    \left[{\bf C}\right]_{IJ} = \delta_{J, c[I]}
\end{equation}
with
\begin{equation}
    c[I] = \left( a\left[\lfloor(I-1) / p\rfloor + 1\right] - 1\right)
    q + b[1 + (I-1)\ \mathrm{mod}\ p].
\end{equation}
Note that in representing arrays we use $1$-based indexing.

%%%%%%%%%%%%%%%%%%%%%%%%%%%%%%%%%%%%%%%%%%%%%%%%%%%%%%%%%%%%%%%%%%%%%%%%%
\subsection*{Direct sum}

The direct sum of $\left[{\bf A}\right]_{p\times q}$ and
$\left[{\bf B}\right]_{r\times s}$ produces the block matrix
$\left[{\bf C}\right]_{(p+r)\times (q+s)}$ where
\begin{equation}
     {\bf C} = \begin{pmatrix}
        {\bf A} & 0\\
        0 & {\bf B}
    \end{pmatrix}.
\end{equation}
The non-zero columns of ${\bf C}$ are the same as those of ${\bf A}$ for the first
$p$ rows. In the next $r$ rows they are the same as those of ${\bf B}$, but
shifted by $q$ spots. In terms of the arrays for these matrices,
\begin{equation}
    c[i] = \begin{cases}
        a[i], & 1 \leq i \leq p,\\
        q + b[i-p], & p < i \leq p+r.
    \end{cases}
\end{equation}
%

%%%%%%%%%%%%%%%%%%%%%%%%%%%%%%%%%%%%%%%%%%%%%%%%%%%%%%%%%%%%%%%%%%%%%%%%%

\subsection*{Matrix concatenation}

To vertically concatenate two row-switching matrices
$\left[{\bf A}\right]_{p\times q}$ and $\left[{\bf B}\right]_{r\times q}$, as in,
\begin{equation}
    \left[{\bf C}\right]_{(p+r)\times q}
    =
    \begin{pmatrix}
        {\bf A} \\ {\bf B}
    \end{pmatrix},
\end{equation}
one simply concatenates their arrays:
\begin{equation}
    c[i] = \begin{cases}
        a[i], & 1 \leq i \leq p,\\
        b[i-p], & p < i \leq p+r.
    \end{cases}.
\end{equation}

The matrix resulting from horizontal concatenation of $\left[{\bf A}\right]_{p\times q}$ and
$\left[{\bf B}\right]_{p\times r}$, 
\begin{equation}
    \left[{\bf C}\right]_{p\times (q+r)} = \begin{pmatrix}
        {\bf A} & {\bf B}
    \end{pmatrix}
\end{equation}
is not row-switching, and displays two non-zero entries per row, 
in distinct blocks. Thus, the matrix ${\bf C}$ can be represented as an array of
ordered pairs,
\begin{equation}
    c[i] = (a[i], b[i])
\end{equation}
for $1\leq i\leq p$.

%%%%%%%%%%%%%%%%%%%%%%%%%%%%%%%%%%%%%%%%%%%%%%%%%%%%%%%%%%%%%%%%%%%%%%%%%
\subsection*{Switch-Unique (SU) decomposition}
\label{SU-decomposition}

\begin{figure}
    \begin{algorithm}[H]
        \caption{SU decomposition using hash tables}
        \label{alg:SU_hashtable}
        \begin{algorithmic}[1] % The number tells where the line numbering should start
            \Procedure{SU-Decompose}{$a$} \Comment{$a$ is an array of numbers or ordered pairs}
            \State $H \gets \text{dictionary that maps keys of same type as } a \text{ to numbers}$
            \State $s \gets \text{array of same length as } a$
            \For {$i \gets 1, \dotsc, \Call{Size}{a}$}
            \State $v \gets a[i]$
            \If {$v \notin \Call{Keys}{H}$}
            \State $H[v] \gets \Call{Size}{H} + 1$ \Comment{insert $v$ into $H$ as a new unique element}
            \EndIf
            \State $s[i] \gets H[v]$
            \EndFor
            \State $u \gets \text{array of length } \Call{Size}{H}$
            \For{$v \gets \Call{Keys}{H}$}
            \State $j \gets H[v]$
            \State $u[j] = v$
            \EndFor
            \State \Return $(s, u)$
            \EndProcedure
        \end{algorithmic}
    \end{algorithm}
\end{figure}

The subroutine implementing the SU decomposition is an important to various BMP operations,
in particular, to CLEAN and SWAP. In both cases, it is performed on
matrices that are concatenated vertically or horizontally. While these
two types of matrices are represented slightly differently, SU
decomposition can be implemented for both using hash tables in a
similar manner.

Let $\left[{\bf A}\right]_{p\times q}$ be the matrix being decomposed,
expressed as an array $a$ of numbers or ordered pairs. We wish to
obtain the array representations of $\left[{\bf A}\right]_{p\times r}$ and
$\left[{\bf U}\right]_{r\times q}$ in
\begin{equation}
    {\bf A} = {\bf S} \; {\bf U},
\end{equation}
which we call $s$ and $u$ respectively, directly from $a$. (Notice
that $s$ is always an array of numbers, while $u$ may be an array of
numbers or ordered pairs depending on the type of $a$.) To this end,
we first create a hash table $H$ which uses the element type of $a$ as
keys and returns numbers. This table is updated while looping over $a$
in such a way that each key is associated with its order of
appearance, i.e., if $v$ is the $i$-th unique element in $a$, $H[v]$
returns $i$. At the end of this loop, we make the assignments
\begin{align}
    s[i] &\gets H[a[i]], \text{for } 1 \leq i \leq p,\\
    u[H[v]] &\gets v, \text{for } v \in H.
\end{align}
See Algorithm \ref{alg:SU_hashtable} for a sketch of this. We note
here that in most cases ${\bf A}$ is a temporary matrix that is not needed
for any purpose other than the SU decomposition, and therefore our
actual implementation uses a slightly less direct approach that
completely avoids its allocation.

%%%%%%%%%%%%%%%%%%%%%%%%%%%%%%%%%%%%%%%%%%%%%%%%%%%%%%%%%%%%%%%%%%%%%%%%%
\acknowledgments

The authors were partially supported by the NSF grants GCR-2428487
(C.C., A.E.R., and U.E.U.) and OIA-2428488 (E.R.M); DOE Grant
DE-FG02-06ER46316; and by a BU Hariri Institute for Computing and
Computational Science \& Engineering Focused Research Group program.

%%%%%%%%%%%%%%%%%%%%%%%%%%%%%%%%%%%%%%%%%%%%%%%%%%%%%%%%%%%%%%%%%%%%%%%%%
\bibliography{references}{}
\bibliographystyle{plain}

%%%%%%%%%%%%%%%%%%%%%%%%%%%%%%%%%%%%%%%%%%%%%%%%%%%%%%%%%%%%%%%%%%%%%%%%%

\end{document}